\documentclass[a4paper,12pt]{article}
\pdfoutput=1
\usepackage{epsfig}
\usepackage{amssymb,subfigure}
\usepackage{amsfonts}
\usepackage{amsmath}
\usepackage{euscript}
\usepackage{verbatim}
\usepackage{latexsym}
\usepackage{graphicx,color}
\usepackage{caption}
\usepackage{float}
\usepackage{slashed}
\usepackage{graphicx}
\graphicspath{ {images/} }
\usepackage{ytableau}

\usepackage{multirow}

\usepackage{mathtools}

\usepackage[colorinlistoftodos]{todonotes}
\usepackage[colorlinks=true, allcolors=blue]{hyperref}
\usepackage{wrapfig}
	\usepackage[T1]{fontenc}
	\usepackage{tikz}
	\usetikzlibrary{decorations.pathmorphing}
\usepackage{tikz}
\usetikzlibrary{shapes.geometric, arrows,patterns,snakes}
\tikzstyle{ellip} = [ellipse, minimum width=3cm, minimum height=1cm,text centered, draw=black]

\newskip\humongous \humongous=0pt plus 1000pt minus 1000pt

\newif\ifdtup

\allowdisplaybreaks[1]

\catcode`\@=11

\@addtoreset{equation}{section}

\def\@normalsize{\@setsize\normalsize{15pt}\xiipt\@xiipt
\abovedisplayskip 14pt plus3pt minus3pt%
\belowdisplayskip \abovedisplayskip
\abovedisplayshortskip \z@ plus3pt%
\belowdisplayshortskip 7pt plus3.5pt minus0pt}

\def\small{\@setsize\small{13.6pt}\xipt\@xipt
\abovedisplayskip 13pt plus3pt minus3pt%
\belowdisplayskip \abovedisplayskip
\abovedisplayshortskip \z@ plus3pt%
\belowdisplayshortskip 7pt plus3.5pt minus0pt
\def\@listi{\parsep 4.5pt plus 2pt minus 1pt
     \itemsep \parsep
     \topsep 9pt plus 3pt minus 3pt}}

\relax

\catcode`@=12

\catcode`\@=11

\def\section{\@startsection{section}{1}{\z@}{3.5ex plus 1ex minus
   .2ex}{2.3ex plus .2ex}{\large\bf}}

\def\SymBoxes#1#2#3#4{\newdimen\un@t \un@t#3%
\raisebox{#1}{\rule{#2\un@t}{#4}\hskip-#2\un@t
\@tempdimb\un@t \advance\@tempdimb by-#4\@tempcntb#2\relax%
\@whilenum{\@tempcntb>0}\do{
\rule{#4}{\un@t}\hskip\@tempdimb \advance\@tempcntb by\m@ne}%
\hskip-#2\un@t \rule[\un@t]{#2\un@t}{#4}%
\rule[\un@t]{#4}{#4}\hskip-#4
\rule{#4}{\un@t}}\hskip-#4}                

\begin{document}

\newcommand{\beq}{\begin{equation}}
\newcommand{\eeq}{\end{equation}}
\newcommand{\bea}{\begin{eqnarray}}
\newcommand{\eea}{\end{eqnarray}}
\newcommand{\beas}{\begin{eqnarray*}}
\newcommand{\eeas}{\end{eqnarray*}}
\newcommand{\defi}{\stackrel{\rm def}{=}}
\newcommand{\non}{\nonumber}
\newcommand{\bquo}{\begin{quote}}
\newcommand{\enqu}{\end{quote}}
\renewcommand{\(}{\begin{equation}}
\renewcommand{\)}{\end{equation}}
\def \eqn#1#2{\begin{equation}#2\label{#1}\end{equation}}
\def\IZ{{\mathbb Z}}
\def\IR{{\mathbb R}}
\def\IC{{\mathbb C}}
\def\IQ{{\mathbb Q}}
\def\de{\partial}
\def\Tr{ \hbox{\rm Tr}}
\def\H{ \hbox{\rm H}}
\def\HE{ \hbox{$\rm H^{even}$}}
\def\HO{ \hbox{$\rm H^{odd}$}}
\def\K{ \hbox{\rm K}}
\def\Im{ \hbox{\rm Im}}
\def\Ker{ \hbox{\rm Ker}}
\def\const{\hbox {\rm const.}}
\def\o{\over}
\def\im{\hbox{\rm Im}}
\def\re{\hbox{\rm Re}}
\def\bra{\langle}\def\ket{\rangle}
\def\Arg{\hbox {\rm Arg}}
\def\Re{\hbox {\rm Re}}
\def\Im{\hbox {\rm Im}}
\def\exo{\hbox {\rm exp}}
\def\diag{\hbox{\rm diag}}
\def\longvert{{\rule[-2mm]{0.1mm}{7mm}}\,}
\def\a{\alpha}
\def\dag{{}^{\dagger}}
\def\tq{{\widetilde q}}
\def\p{{}^{\prime}}
\def\W{W}
\def\N{{\cal N}}
\def\hsp{,\hspace{.7cm}}

\def\br{\nonumber\\}
\def\IZ{{\mathbb Z}}
\def\IR{{\mathbb R}}
\def\IC{{\mathbb C}}
\def\IQ{{\mathbb Q}}
\def\IP{{\mathbb P}}
\def \eqn#1#2{\begin{equation}#2\label{#1}\end{equation}}

\newcommand{\sgm}[1]{\sigma_{#1}}
\newcommand{\idd}{\mathbf{1}}

\newcommand{\C}{\ensuremath{\mathbb C}}
\newcommand{\Z}{\ensuremath{\mathbb Z}}
\newcommand{\R}{\ensuremath{\mathbb R}}
\newcommand{\rp}{\ensuremath{\mathbb {RP}}}
\newcommand{\cp}{\ensuremath{\mathbb {CP}}}
\newcommand{\vac}{\ensuremath{|0\rangle}}
\newcommand{\vact}{\ensuremath{|00\rangle}                    }
\newcommand{\oc}{\ensuremath{\overline{c}}}
\begin{titlepage}
\begin{flushright}
CHEP XXXXX
\end{flushright}
\bigskip
\def\thefootnote{\fnsymbol{footnote}}

\begin{center}
{\large
{\bf 
Interpreting the Bulk Page Curve: \\ 
\vspace{0.1in}
{\large A Vestige of Locality on Holographic Screens}
}
}
\end{center}

\bigskip
\begin{center}
{ Chethan KRISHNAN$^a$\footnote{\texttt{chethan.krishnan@gmail.com}}, \ \  Vyshnav MOHAN$^a$,\footnote{\texttt{vyshnav.vijay.mohan@gmail.com}} \ \ \vspace{0.15in} \\ }
\vspace{0.1in}

\end{center}

\renewcommand{\thefootnote}{\arabic{footnote}}

\begin{center}

$^a$ {Center for High Energy Physics,\\
Indian Institute of Science, Bangalore 560012, India}

\end{center}

\noindent
\begin{center} {\bf Abstract} \end{center}
Areas of extremal surfaces anchored to sub-regions on screens in Minkowski space satisfy various entanglement entropy inequalities. In 2+1 dimensions where the arguments are simplest, we demonstrate (a) monogamy of mutual information, (b) various versions of (strong) subadditivity, (c) various inequalities involving the  entanglement of purification, as well as (e) reflection inequality and (f) Araki-Lieb inequality. Just as in AdS, Linden-Winter and the tower of Cadney-Linden-Winter inequalities are satisfied trivially. All of these are purely geometric (and therefore unambiguous) statements, and we expect them to hold semi-classically when $G_N \rightarrow 0$. The results of arXiv:2103.17253 suggest that it is unlikely that there is non-analyticity at $G_N=0$. These observations have relevance for the Page phase transition in flat space black holes observed with respect to a screen in arXiv:2005.02993 and arXiv:2006.06872. In particular, they constitute a Lorentzian argument that these extremal surface transitions are indeed phase transitions of {\em some} suitably defined entanglement entropy associated to {\em subregions} on the screen.


\vspace{1.6 cm}
\vfill

\end{titlepage}

\setcounter{page}{2}

\setcounter{footnote}{0}



\section{Context}

The island idea \cite{Penington, Almheiri} is a recent and thought-provoking step forward, in our efforts to resolve the black hole information paradox \cite{Hawking, Page, Mathur, AMPS}. It has presented us with a natural way to think about the unitary Page curve for black holes in terms of entanglement entropy, geometrized via semi-classical Ryu-Takayanagi (RT) surfaces \cite{RT} and their generalizations \cite{HRT,EW}. 

These calculations are cleanest for AdS black holes  coupled to sinks/baths, where the entanglement entropy has a clear understanding in terms of subregions on the AdS boundary. However, it was noticed fairly quickly \cite{Jude, Maldacena} that they allow a very natural generalization to flat space black holes as well. If one defines (quantum) extremal/RT surfaces with respect to a suitably defined screen around the black hole, areas of these surfaces again undergo a phase transition at the Page time, leading to a tent-shaped Page curve. In 1+1 dimensions, where the discussion is somewhat special for multiple reasons\footnote{The most crucial one for our purposes here will be that there are no subregions in 1+1 dimensions.}, related observations had appeared even earlier \cite{Thorlacius}.


In AdS, the RT surface prescription for computing entanglement entropy has been thoroughly vetted, due largely to the fact that we know the dual theory -- it is a conventional {\em local} quantum field theory. Subregions on the boundary therefore have natural interpretations as tensor factors. On the other hand, it is less clear {\em what} the extremal surfaces are computing, in flat space. Evidence presented in \cite{ACD, Jude}, that screen subregions share some of the crucial properties of AdS boundary subregions, relied on objects called Asymptotic Causal Diamonds (ACDs). These are causal diamonds whose vertices are attached to the null boundaries of asymptotically flat space. It was noted that these objects, despite the fact that the conformal boundary of flat space has no well-defined causal structure, can serve as a well-defined kinematic space in the spirit of \cite{Czech}. It was noted earlier in \cite{ACD} that ACDs together with screens can reproduce quantum error correction like features in flat space. Causal structure based arguments and ACD constructions were used to prove theorems like strong sub-additivity for maxi-min surfaces defined on screens \cite{Jude}. The naturalness with which ACDs and screens worked together to give HRT-like \cite{HRT} and maximin-based \cite{Wall} definitions of extremal surfaces was one of the major motivations for the proposals in \cite{Jude}. Another motivation for associating entanglement entropy to screens in \cite{Jude} came from Euclidean arguments going back to Gibbons and Hawking \cite{GH}, where also a cut-off played a crucial role\footnote{The Gibbons-Hawking argument was emphasized in \cite{Maldacena} as well.}. Based on all these motivations, in \cite{Jude}, a proposal was made for the generalized entropy of sub-regions of the screen with the exterior region viewed as a sink. 

ACDs in flat space are not widely known to the community beyond the authors of \cite{ACD, Jude}. Besides, the ACD based theorems of \cite{Jude} are somewhat technical and rely on causal structure based arguments in general relativity. In this paper, we will present an extremely elementary set of observations to give strong evidence that areas of extremal surfaces anchored to subregions on screens in Minkowski space, do behave like entanglement entropy. The idea is simply that entanglement entropy is well-known to satisfy a large number of inequalities. So we can check these inequalities for subregions on screens. This will be an argument for believing that these extremal surface areas should really be viewed as entanglement entropies\footnote{If it looks like a duck, walks like a duck and quacks like a duck, ...}.

A further key feature of our observations is that they can be phrased entirely in {\em Lorentzian} signature, in terms of screens in Minkowski space. In that sense, they are complementary to the Euclidean arguments in \cite{Jude, Maldacena}, and they are a poor man's version of the ACD arguments that we found compelling in \cite{Jude}. Another  comment worth making is that since we are working with simple geometric statements in Minkowski space, we can go much beyond the strong sub-additivity theorem we proved using ACDs in {\em asymptotically} Minkowski space. We can prove {\em all} the entanglement inequalities satisfied by AdS subregions (as far as we are aware), for flat space screens as well. Just as is done usually in AdS, to avoid geometric complications, we will work in this paper with 2+1 dimensions. We will find that the proofs are even more simpler than they were in AdS, and have ordinary triangle inequality as a key ingredient. 

In the next few sections, we will prove various versions of (strong) sub-additivity, monogamy of mutual information, various inequalities involving the entanglement of purification, reflection inequality, as well as Araki-Lieb, Linden-Winter and Cadney-Linden-Winter inequalities. The inequalities come with various subcases, which we prove as we go. We also extract a general principle which enables these proofs (and has to do with the entanglement wedges of subregions), in an appendix. In a final section, we discuss what the broader lessons are for the Page curve for black holes, arising from these observations. The reader willing to believe our elementary proofs may want to skip directly to this section.

{\bf Note added:} The present paper contains results that are more than two years old. They are an extension of the main results we have already presented in \cite{ACD} and \cite{Jude}. But multiple interactions\footnote{\label{foot}Including questions from a JHEP referee of \cite{Kausik} to the effect: what is the big deal about some minimal surface phase transitions?} during the last year have convinced us that some of the implications of these results have not penetrated the zeitgeist, specifically in the context of the recent discussions on the flat space bulk Page curve. A key message is that there {\em should} exist a  notion of entanglement entropy, that one can associate to {\em subregions} of suitable screens in flat space, even though at this point in history it is not particularly clear how to do this from first principles\footnote{The ACD based arguments make it plausible that some fuzziness in the time direction on the screen may play a role, but not enough work has been done to develop this.}. A suggestion that follows from \cite{Jude, Vyshnav, Kausik} is that the screen size should be viewed as a scale of coarse-graining. It seems also possible that since screens deal with bulk local quantities, demanding full gauge invariance may not be necessary. Whatever way these questions eventually play out, the observations remain. They also strengthen the case for a conventional bulk Page curve in flat space as argued in \cite{Jude, Maldacena}. Since this has been contested in \cite{Alok, Andreas} which appeared after \cite{ACD}, we feel that it is appropriate to report these observations here. 


\section{Entanglement Entropy and Holographic Screens}

\begin{figure}[t]
\centering
\includegraphics[scale=0.45]{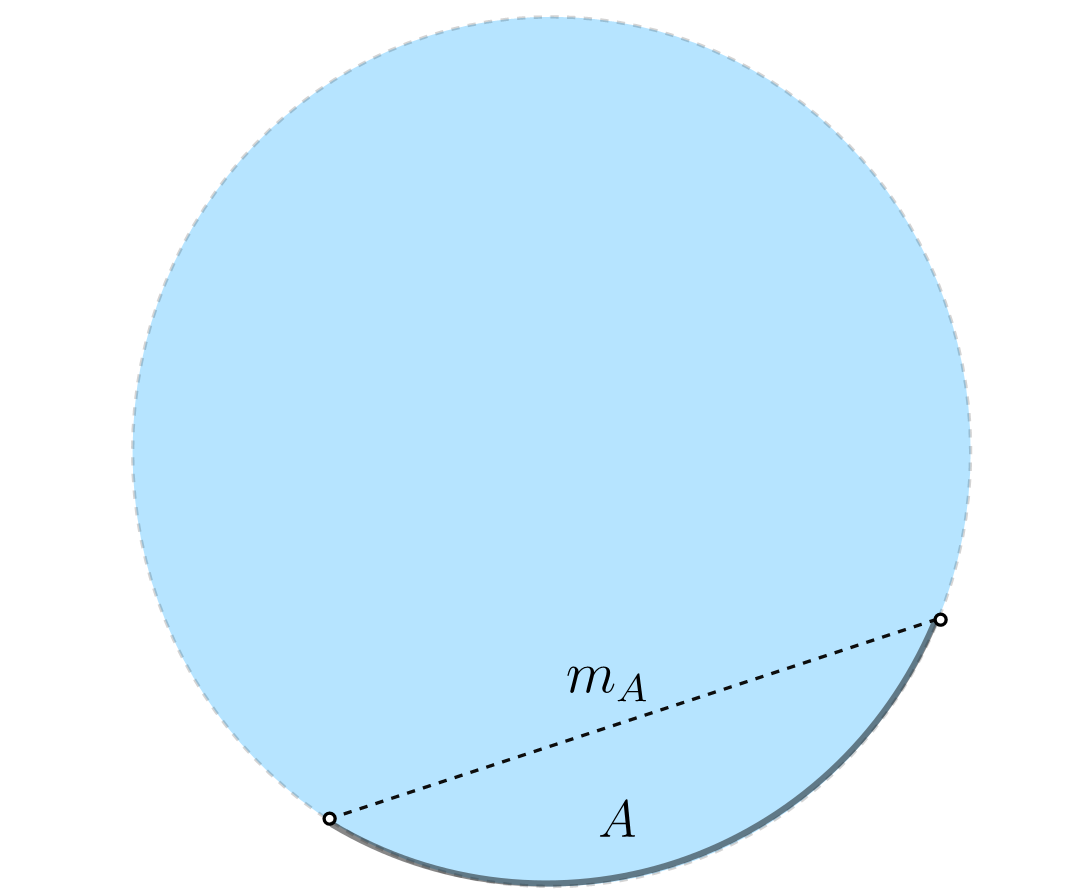}
\caption{Consider flat-space with a radial cutoff. This region is represented by the blue disk. A region $A$ on the cutoff surface is denoted by the gray arc in the above diagram. The minimal surface of this region is the straight line joining the endpoints of the arc. We will denote the length of this minimal surface by $m_A$.}
\label{rtfig}
\end{figure}

The underlying idea that we wish to emphasize in this paper is that we can associate an entanglement entropy to subregions on screens in Minkowski space. A cross section of Minkowski space within a timelike cylinder of fixed raius at a fixed instant is shown in Figure \ref{rtfig}, and a subregion is denoted by $A$. The obvious claim, inspired by the well-known Ryu-Takayanagi prescription is that one can associate an entanglement entropy to $A$ via the area 
\begin{equation}
S(A)=\frac{1}{4 G_{\mathrm{N}}} \  \min _{m: \partial m=\partial A} a(m) \label{rteq}
\end{equation}
where $m$ is a hypersurface in the bulk that ends on $\partial A$ and $a(m)$ is the area of this hypersurface. The hypersurfaces have an additional constraint that they should be  homologous to $A$, i.e., there should exist a region $r$ in the bulk such that $\partial r =A \cup m$. All the regions and surfaces that we will consider are at a constant-time slice of the spacetime. We will refer to these surfaces as \textit{minimal surfaces}. In Minkoswki space, these are just hyperplanes and as is customary in AdS we will work with 2+1 dimensions for technical simplicity. Our results lead us to believe that the picture that emerges is isomorphic for AdS and flat space, so we expect that analogues of AdS statements will hold in higher dimensional flat space as well. 

Some of our theorems are most conveniently phrased in terms of \textit{mutual information}, a quantity built from entanglement entropy that remains finite when the UV cutoff the theory is taken to infinity. For disjoint regions, it is given by
\begin{equation}
I(A:B) = S(A) +S(B) -  S(A \cup B).
\end{equation}
Another such quantity that will be useful for us in the upcoming discussions is called \textit{conditional mutual information}, defined by
\begin{equation}
I(A: C | B):=S(A \cup  B)+S(B \cup C)-S(A \cup B \cup C)-S(B)=I(A: B \cup C)-I(A: B)
\end{equation}

Our goal in most of the rest of this paper is to prove various theorems that are known to be satisfied by entanglement entropy for the quantity $S(A)$, thereby building evidence that this quantity is indeed an entanglement entropy\footnote{Note that we are agnostic about the precise definition of this entanglement entropy in this paper. But a definition in terms of a split of the full spacetime into a holographic system inside the screen coupled to a sink outside, has been proposed in section 4.4 of \cite{Jude}, when gravity is weakly turned on. Our discussion here should be compared to the exercise of computing RT surfaces in empty AdS. But we wish to phrase the statements in this paper as purely geometric statements  in a fixed spacetime, with no reference to any theory. Therefore the $G_{\rm N}$ in \eqref{rteq} is simply a book keeping constant as far as we are conserved.}. 

We apply the prescription in \eqref{rteq} to $2+1$ dimensional Minkowski space with a radial cutoff. The minimal surfaces on a constant-time slice in this space turns out to be straight lines. Thus, to calculate the entanglement entropy of a region $A$ on the cutoff surface, it suffices to find the length of the straight line which ends on $\partial A$ and is homologous to $A$ (See figure \ref{rtfig}). This simplicity makes our calculations considerably technically simpler in 2+1 than in higher dimensions -- analogous facts are routinely exploited in AdS also where much of the work is done in AdS$_3$. We will refer to papers on entanglement entropy inequalities in AdS as
relevant -- an original reference is \cite{Headrick}, see also various discussions in eg., \cite{entAdS}. 

\subsection{Inequalities}

In this subsection, we list a number of inequalities that are known to be satisfied by entanglement entropy. We will prove them for the quantity $S(A)$ we defined in the last section. We will also state and prove more entanglement inequalities (involving an auxiliary quantity called the "entanglement of purification"), in a later section.

In a \textit{general} quantum system, the entanglement entropy of its subsystems satisfies the following inequalities: 
\begin{enumerate}
\item \textit{Subadditivity} : $S(A)+S(B) \geq S(A \cup B)$
\item \textit{Araki-Lieb} : $S(A \cup B) \geq \left|S(A)-S(B)\right|$
\item \textit{Strong Subadditivity I} : $S(A \cup B)+S(B \cup C) \geq S(A \cup B \cup C)+S(B)$
\item \textit{Strong Subadditivity II} : $S(A \cup B)+S(B \cup C) \geq S(A)+S(C)$.
\end{enumerate}
Entanglement entropy also satisfies other constrained inequalities that are best expressed in terms of the conditional mutual information. In terms of this quantity, we have:
\begin{enumerate}
\setcounter{enumi}{4}
\item \textit{Linden-Winter} :  if 
\begin{equation}
I(A: C | B) = I(A: B | C) =I(B: C | D)=0
\)
then $I(C: D) \geq I(C: A \cup B)$
\item \textit{Cadney-Linden-Winter} : if $I(A: C | B)  = I(B: C | A) = 0$, then 
\(
S\left(X_{1} \ldots X_{n}\right)+(n-1) I(A B: C) \leq \sum_{i=1}^{n} S\left(X_{i}\right)+\sum_{i=1}^{n} I\left(A: B | X_{i}\right)
\)
for disjoint subsystems $\{ A,B,X_{1},X_{2},...,X_{n}\}$.
\end{enumerate}
There also exists inequalities that are not satisfied by a generic quantum system but are conjectured to be a constraint on a theory that has a holographic dual \cite{Hayden}. The most important of these is the \textit{monogamy} property of entanglement entropy :
\begin{enumerate}
\setcounter{enumi}{6}
\item \textit{Monogamy} : $S(A\cup B)+S(A \cup C)+S(B \cup C) \geq S(A \cup B)+S(A)+S(B) +S(C)$.
\end{enumerate}
In terms of mutual information, this property can be written as
\(
I(A:B)+I(B:C) \leq I(A:B \cup C).
\)
In addition to the above inequalities, a new property called the refection inequality was proposed in \cite{Casini:2010nn, Headrick:2013zda}. Consider two boundary regions $A$ and $B$. If we denote the reflection of these regions about an axis by $\bar{A}$ and $\bar{B}$ respectively, we can state the property as follows:
\begin{enumerate}
\setcounter{enumi}{7}
\item \textit{Reflection} : $S(A \cup \bar{A})+S(B \cup \bar{B}) \leq S(A \cup \bar{B})+S(B \cup \bar{A}) $.
\end{enumerate}

We will show in the following (sub-)sections that the prescription for flat space screens satisfies all of these inequalities. But before we get to the proofs, let us briefly discuss what the minimal surfaces for various regions are. 

\subsection{Minimal Surfaces}

For regions which are continuous on the boundary, the minimal surface is the straight line joining the end points (See figure \ref{rtfig}). The same picture can be extended to the union of regions that are contiguous (See figure \ref{subadd}).

\begin{figure}[!tbp]
  \centering
  \begin{minipage}[b]{0.48\textwidth}
    \includegraphics[scale=0.45]{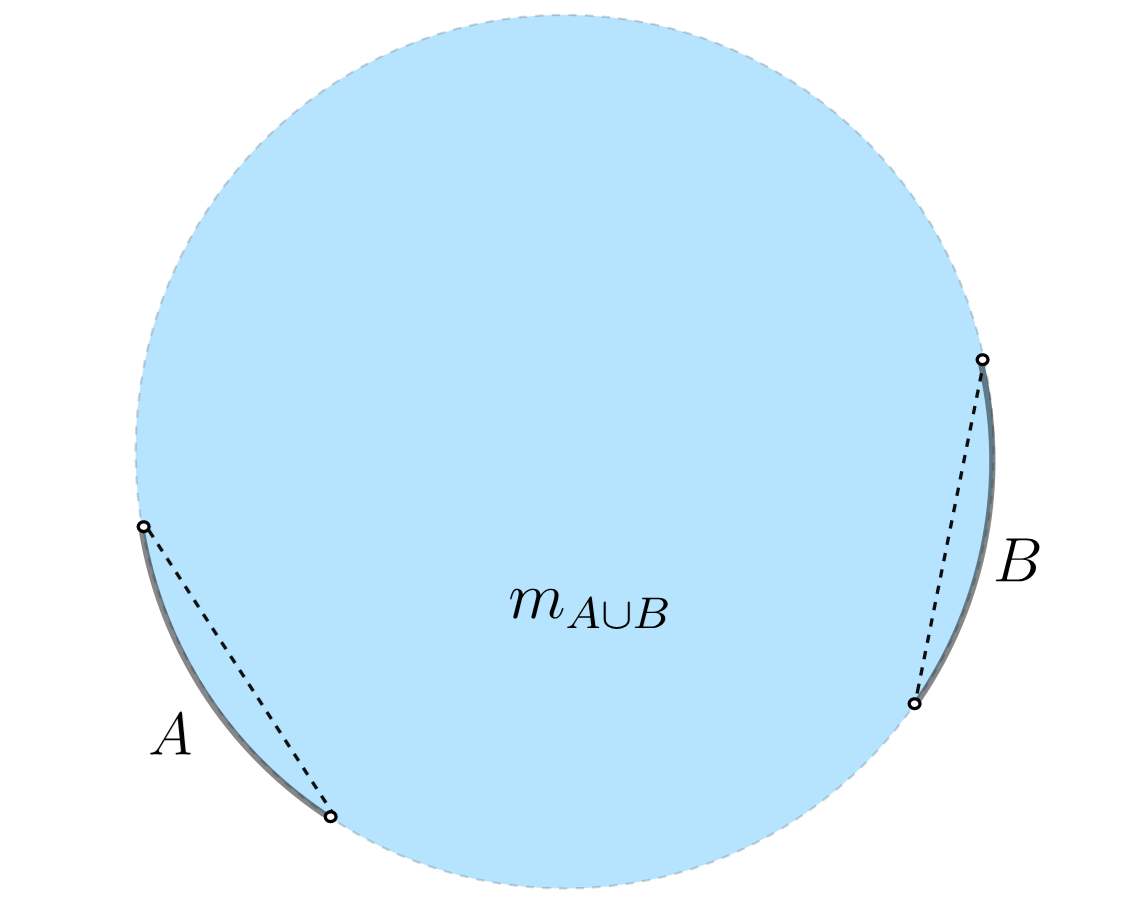}
\begin{center}
 (a)
\end{center}
  \end{minipage}
  \hfill
  \begin{minipage}[b]{0.48\textwidth}
    \includegraphics[scale=0.45]{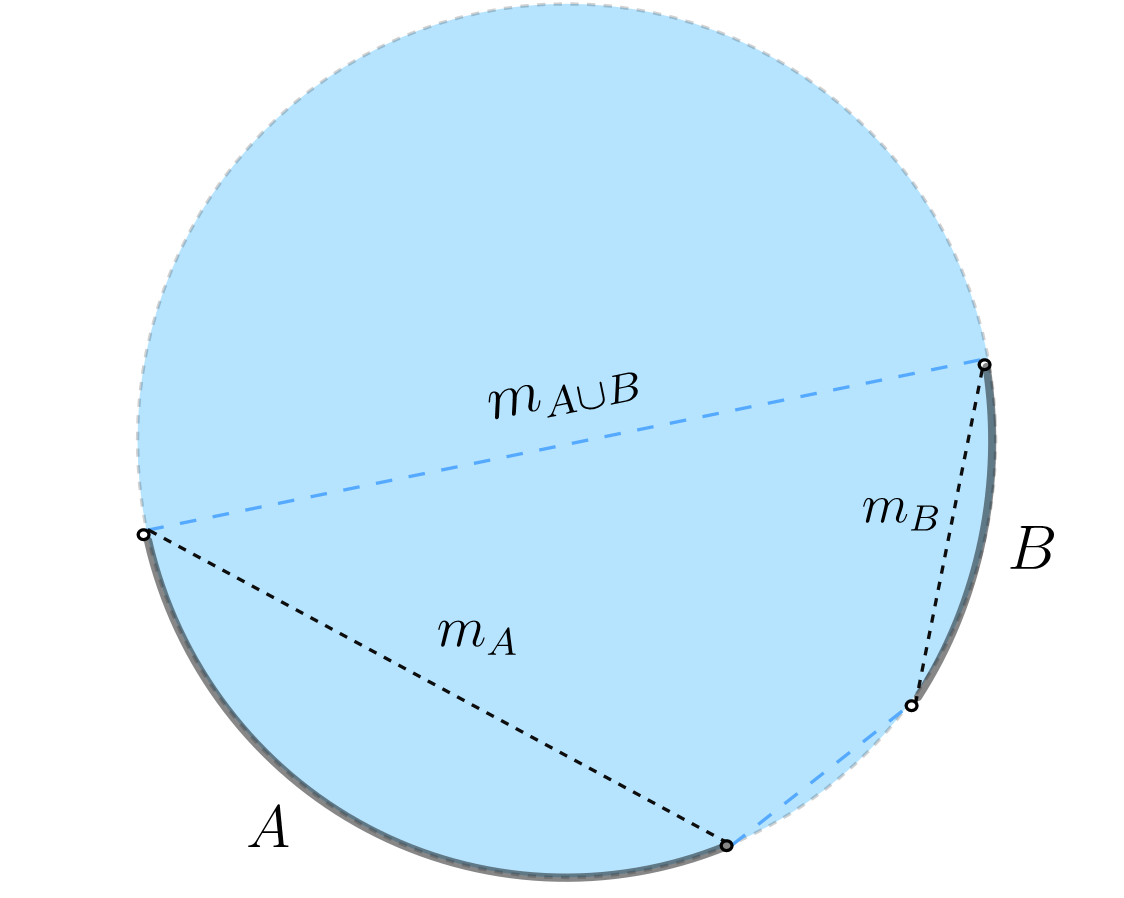}
\begin{center}
 (b)
\end{center}
  \end{minipage}
\caption{The minimal surfaces of union of regions that are not adjacent to each other \textbf{a.} When the separation between the regions is much larger than the size of the region, the minimal surface of $A \cup B$, is the union of the minimal surfaces of $A$ and $B$. These surfaces are denoted by the dotted lines. \textbf{b.} When the separation between the two regions is much smaller then the size of the regions, the minimal surface is given by union of the blue dotted lines in the figure.}
\label{minimalfig2}
\end{figure}

However, for regions that are not adjacent to each other, there could be many surfaces that satisfy the homology constraint (See figure \ref{minimalfig2}). The entanglement entropy will be given by the surface with the minimal length. In particular, let us look at the union of two disjoint boundary regions. It is easy to that when the separation between the regions is very large compared to the size of the regions, figure \ref{minimalfig2}(a) will be the minimal surface. Since the resulting entanglement wedge is disconnected, we will refer to these surfaces as disconnected minimal surfaces. The mutual information vanishes as we have
\(
S(A)+S(B) = S(A \cup B) \label{minimal1}
\)
When the separation of the regions is much smaller than the size of the regions, figure \ref{minimalfig2}(b) will be the minimal surface. Since the corresponding entanglement wedge is connected, we will refer to these surfaces as connected minimal surfaces. We have

\(
S(A)+S(B) \geq S(A \cup B) \label{minimal2}
\)
Therefore, the minimal surfaces undergo a phase transition when we vary the size and separation of the regions.

\section{Proofs}

In rest of the paper, we will denote the length of the minimal surface of a region $A$ by $m_{A}$. We will also use the notation $PQ$ to denote the length of a line segment with endpoints $P$ and $Q$. The proofs are shockingly simple\footnote{Even more so than in AdS.}, and one is left wondering if triangle inequality is the ultimate entanglement inequality. 


\subsection{Subadditivity}
\begin{figure}[t]
\centering
\includegraphics[scale=0.45]{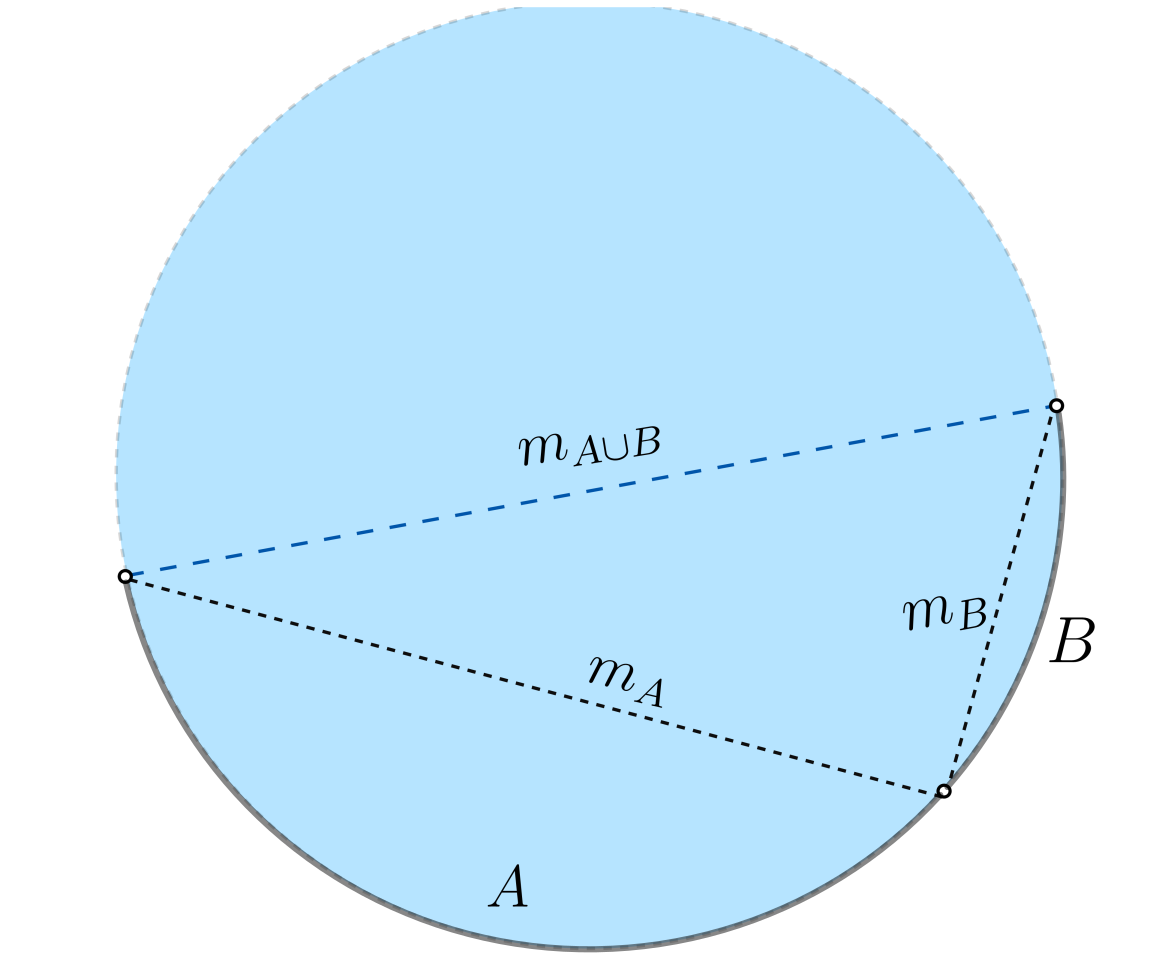}
\caption{The regions $A$ and $B$, denoted by the gray arc, has the minimal surfaces indicated by the black dotted lines. These surfaces have the lengths $m_A$ and $m_B$ respectively. The minimal surface of $A \cup B$ is denoted by the blue dashed line and it has a length of $m_{A \cup B}$.}
\label{subadd}
\end{figure}
\subsubsection{For contiguous regions}
Consider two regions $A$ and $B$ as shown in figure \ref{subadd}. By triangle inequality, we have
\begin{equation}
m_{A} + m_{B} \geq m_{A \cup B}
\end{equation}
Using \eqref{rteq}, we get the subadditivity condition
\(
S(A)+S(B) \geq S(A \cup B)
\)
\subsubsection{For non-contiguous regions}
When the regions are not adjacent to each other, the minimal surface of $A \cup B$ can be either the surface shown in figure \ref{minimalfig2}(a) or \ref{minimalfig2}(b). From \eqref{minimal1} and \eqref{minimal2}, we can see that both these cases satisfy the inequality trivially.
\subsection{Araki-Lieb inequality}
\subsubsection{For contiguous regions}
Consider two regions as in figure \ref{subadd}. By reverse triangle inequality, we have
\begin{equation}
m_{A \cup B} \geq \left| m_{A} - m_{B}\right|
\end{equation}
Dividing by $4 G_{\mathrm{N}}$, we get the required inequality.
\(
S(A \cup B) \geq \left|S(A)-S(B)\right|
\)
\subsubsection{For non-contiguous regions}
There are two potential minimal surfaces when the regions are non-contiguous. When the minimal surface is as in figure \ref{minimalfig2}(a), we have
\(
 S(A \cup B) =S(A)+S(B) \geq  \left|S(A)-S(B)\right|
\)
Therefore, the inequality is satisfied. Now let us look at the case where the minimal surface is as in figure \ref{minimalfig2}(b). Let us assume that $m_{A} \geq m_{B}$ without the loss of any generality. The sum $m_{A \cup B} +m_{B} $ corresponds to the length of a curve that connects the endpoints of $A$. This sum will always be greater than the length of the minimal curve of this region, i.e., $m_{A}$. This gives us
\(
m_{A \cup B} +m_{B} \geq m_{A} \implies m_{A \cup B} \geq m_{A} - m_{B} =  \left| m_{A} - m_{B}\right|
\)
Thus, the Araki-Lieb inequality is satisfied in all the cases.
\subsection{Strong subadditivity I}
\subsubsection{For contiguous regions}
\begin{figure}[t]
  \centering
  \begin{minipage}[b]{0.48\textwidth}
    \includegraphics[scale=0.41]{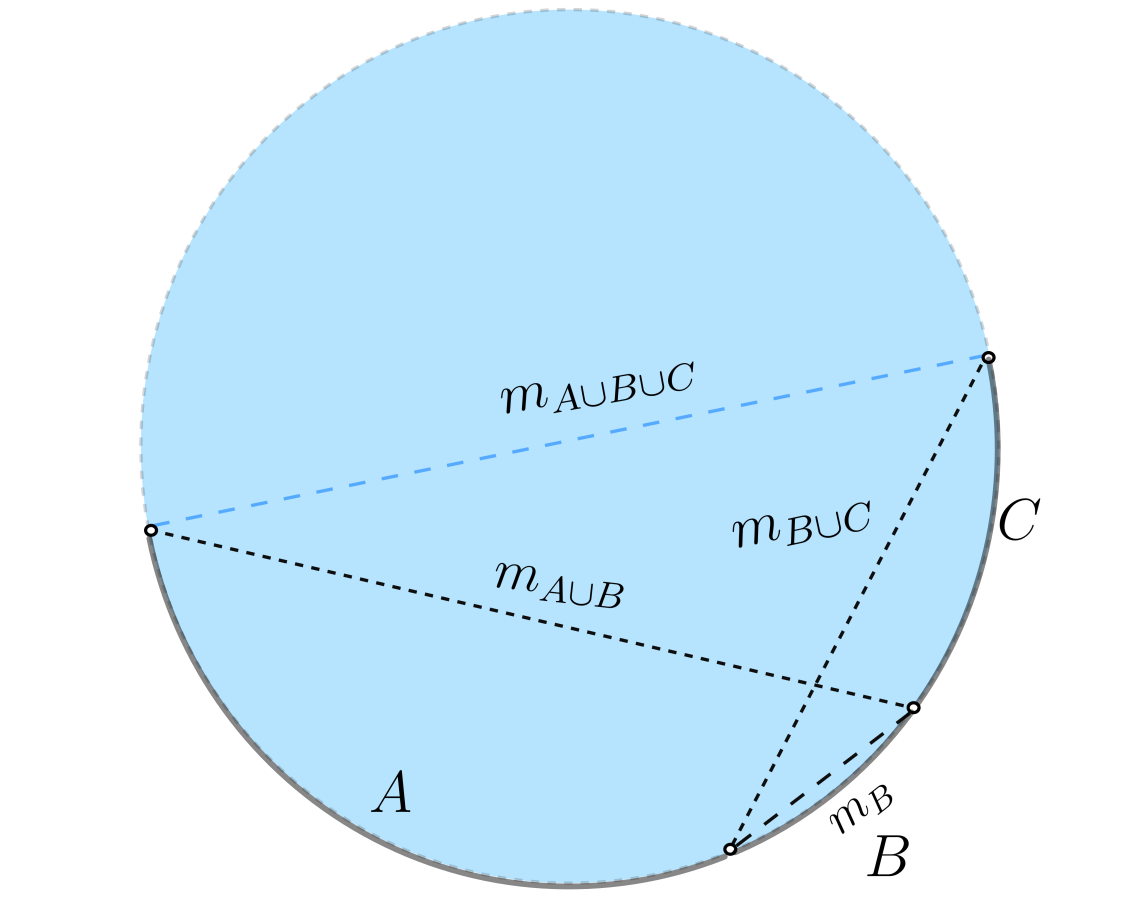}
\begin{center}
 (a)
\end{center}
  \end{minipage}
  \hfill
  \begin{minipage}[b]{0.48\textwidth}
    \includegraphics[scale=0.41]{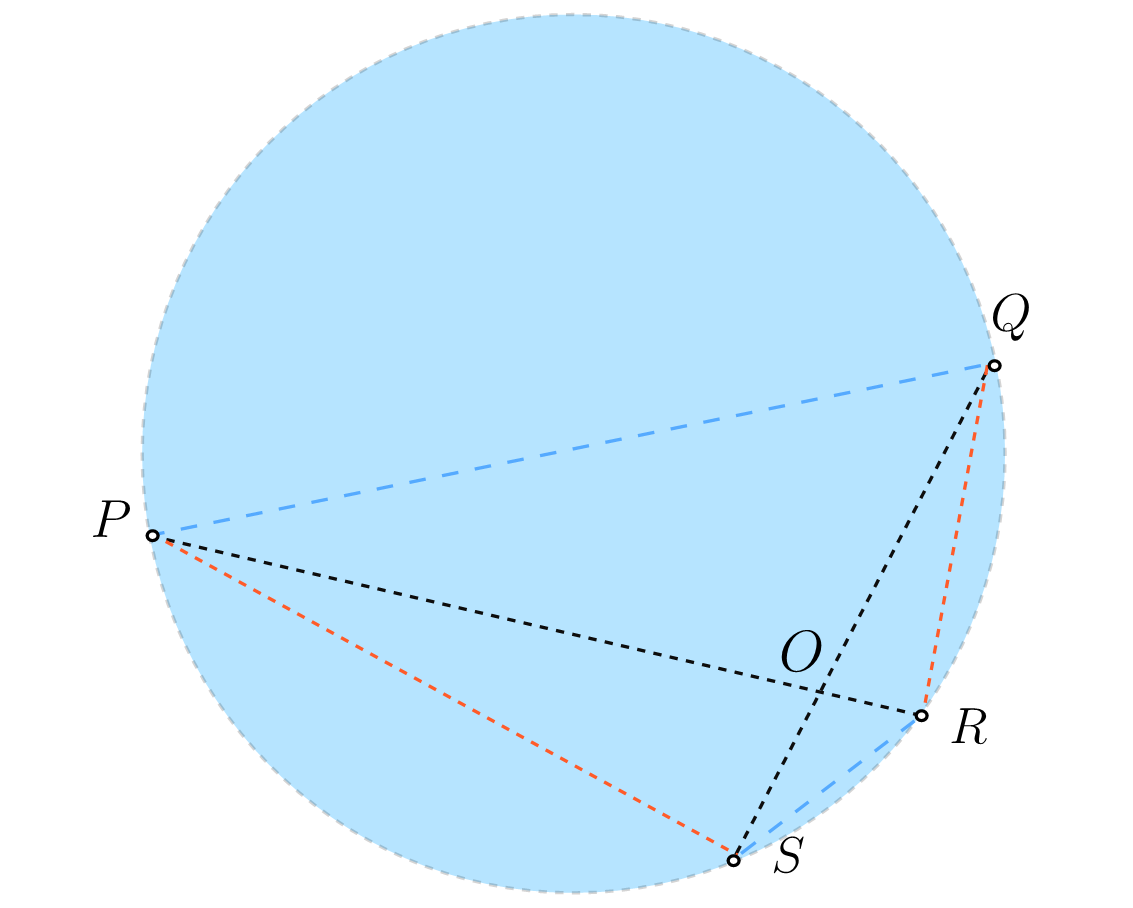}
\begin{center}
 (b)
\end{center}
  \end{minipage}
\caption{\textbf{a.} For regions $A$, $B$ and $C$ that are adjacent to each other, the minimal surface of $A \cup B \cup C$ is indicated by the blue dashed lines. The minimal surfaces of $A \cup B$ and $B \cup C$ are denoted by black dotted lines. \textbf{b.} Representation of all the relevant surfaces using their endpoints on the cutoff surface.}
\label{strongsubfig1}
\end{figure}
Let us consider three adjacent regions $A$, $B$ and $C$ as in figure \ref{strongsubfig1}(a). Keeping only the required surfaces and naming the vertices $P$, $Q$, $R$, $S$ and $O$, we obtain figure \ref{strongsubfig1}(b). We have
\(
PR +QS = PO+OR+QO +OS
\)
From 
\(
 PO+QO \geq PQ \ \ \ \text{and} \ \ \ OR+OS \geq RS \label{triangsubadd}
\)
we have
\(
m_{A \cup B} + m_{B \cup C} = PR +QS \geq PQ+ RS = m_{A \cup B \cup C}+m_{B}
\)
Therefore, we have the inequality
\(
S(A \cup B)+S(B \cup C) \geq S(A \cup B \cup C)+S(B)
\)
In appendix \ref{subaddappend}, we will discuss another proof that relies on a \textit{local surgery} argument.
\subsubsection{For non-contiguous regions}
\begin{figure}[t]
  \centering
  \begin{minipage}[b]{0.48\textwidth}
    \includegraphics[scale=0.41]{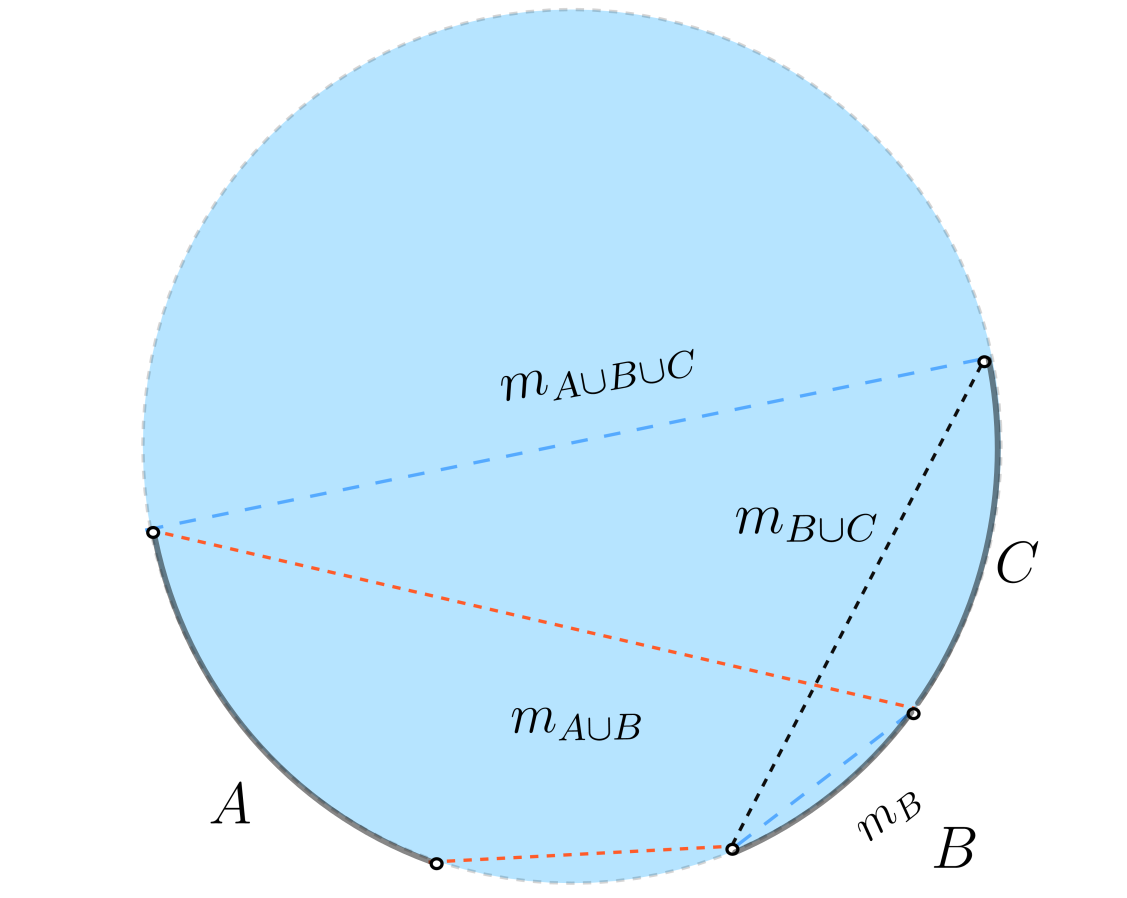}
\begin{center}
 (a)
\end{center}
  \end{minipage}
  \hfill
  \begin{minipage}[b]{0.48\textwidth}
    \includegraphics[scale=0.41]{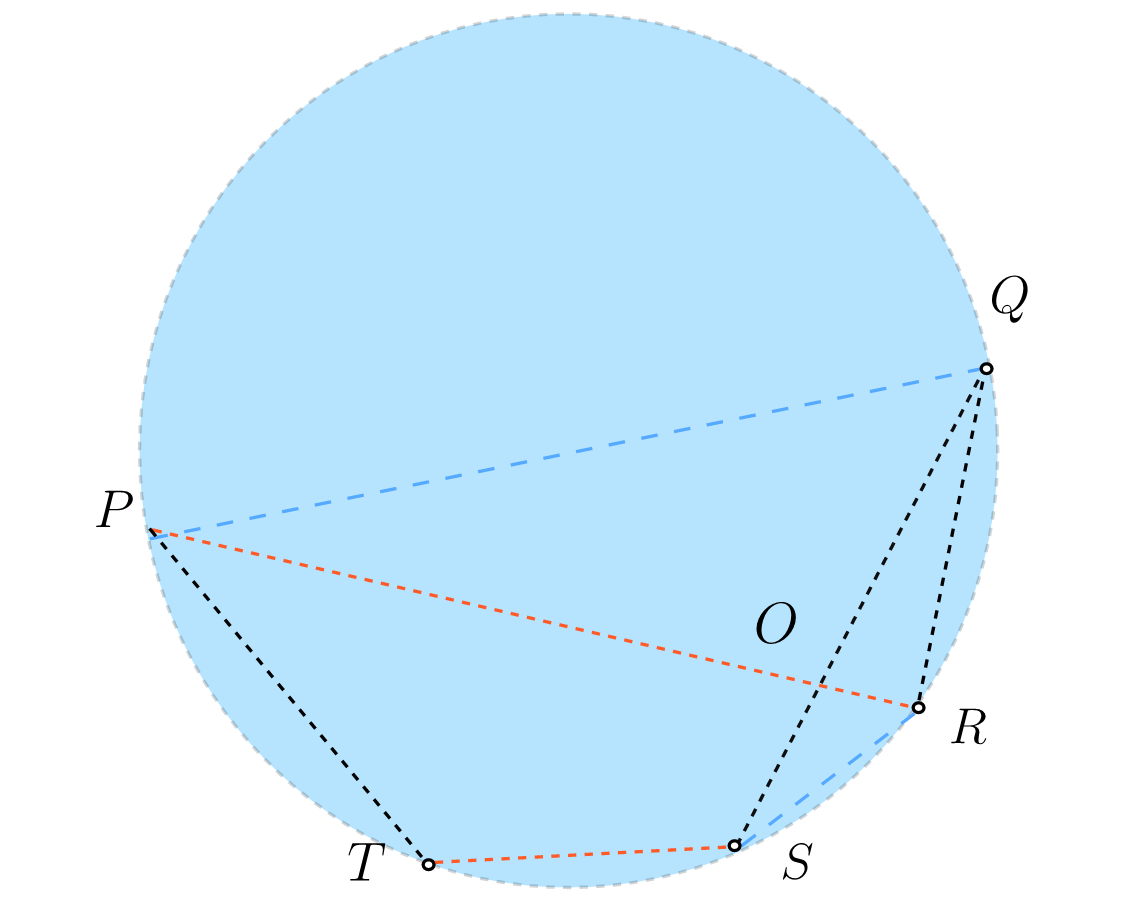}
\begin{center}
 (b)
\end{center}
  \end{minipage}
\caption{\textbf{a.} We consider regions $A$, $B$ and $C$ as shown in the figure. $C$ is adjacent to $B$ while $A$ is not. Compared to the minimal surfaces in figure \ref{strongsubfig1}, there will be an additional term in $m_{A \cup B}$  and $m_{A \cup B \cup C}$ coming from the length of the bottom red dotted line. \textbf{b.} Representation of all the relevant surfaces using their endpoints on the cutoff surface. }
\label{strongsubfig2}
\end{figure}
When the regions $A$, $B$ and $C$ have disconnected minimal surfaces, the inequality saturates; 
\(
S(A \cup B)+S(B \cup C) = S(A) + S(B)+S(B) +S(C) = S(A \cup B \cup C) +S(B)
\)
Now let us consider the case where $A$ is not adjacent to $B$, while $C$ is (See figure \ref{strongsubfig2}). We can see that
\(
PR +QS +TS= PO+OR+QO +OS +TS
\)
From
\(
 PO+QO \geq PQ \ \ \ \text{and} \ \ \ OR+OS \geq RS \label{triangsubadd}
\)
we get
\(
m_{A \cup B} + m_{B \cup C} = (PR+TS) +QS \geq (PQ+TS)+ RS =  m_{A \cup B \cup C}+m_{B}
\)
By transposing the regions, we can prove the inequality in all the other cases as well.
\subsection{Strong subadditivity II}
Proceeding as in the previous section, we can see from from figure \ref{strongsubfig1}(b) that
\(
PR +QS = PO+OR+QO +OS
\)
Using 
\(
 PO+OS \geq PS \ \ \ \text{and} \ \ \ OR+OQ\geq QR
\)
we get,
\(
m_{A \cup B} + m_{B \cup C} = PR +QS \geq PS+QR = m_{A}+m_{C}
\)
This gives us the relation
\(
S(A \cup B)+S(B \cup C) \geq S(A)+S(C)
\)
\subsection{Monogamy}

\subsubsection{For contiguous regions}
Let us consider the regions $A$, $B$ and $C$ as in figure \ref{strongsubfig1}. From
\(
 PO+OS \geq PS \ \ \ \text{and} \ \ \ OR+OQ\geq QR, \label{triangsubadd1}
\)
we get
\(
\begin{aligned}
m_{A \cup B} + m_{B \cup C} + m_{A \cup C} &= PR+QS+ ( PQ +RS)  \\
& = PO+OR+QO +OS+PQ + RS\\
&\geq PS+ RS+QR +PQ = m_{A}+m_{B} +m_{C}+m_{A \cup B \cup C} 
\end{aligned}
\)
Therefore, we have inequality
\(
S(A\cup B)+S(A \cup C)+S(B \cup C) \geq S(A \cup B \cup C)+S(A)+S(B) +S(C)
\)
\subsubsection{For non-contiguous regions}
Let us focus on the case where $A$ is not adjacent to $B$ and $C$, as shown in figure \ref{strongsubfig2}. We have
\(
\begin{aligned}
m_{A \cup B} + m_{B \cup C} + m_{A \cup C} &= (PR+TS)+QS+ ( PQ +RT)  \\
& = PO+OR+QO +OS+TS+ PQ + RT\\
&\geq (PS+QR)  +TS+ PQ + RT \ \ \ \ \  \ \ \ \ \ \left[\text{From} \  \eqref{triangsubadd1}\right]
\end{aligned}
\)
Let us denote $M$ as the intersection of the two lines whose lengths are $PS$ and $RT$. From 
\(
 PM+MT \geq PT \ \ \ \text{and} \ \ \ MR+MS \geq SR
\)
we have 
\(
\begin{aligned}
PS+QR  +TS+ PQ + RT  &= (PM+MS)+QR  +TS+ PQ + (RM+MT) \\
&\geq PT+SR +QR  +TS+ PQ \\
&= m_{A}+m_{B} +m_{C}+m_{A \cup B \cup C}
\end{aligned}
\)
This immediately translates to the monogamy relation. When the regions are non-adjacent, the same proof can be extended trivially. 

When all the regions have a disconnected minimal surface, we have
\(
\begin{aligned}
S(A\cup B)+S(A \cup C)+S(B \cup C)  &= S(A)+S(B)+S(A)+S(C)+S(B) +S(C) \\
& = (S(A)+S(B) +S(C)) + S(A)+S(B) +S(C) \\
&= S(A \cup B \cup C)+S(A)+S(B) +S(C)
\end{aligned}
\)
Therefore, the inequality saturates.
\subsection{Linden-Winter}
We saw that strong subadditivity is saturated, i.e. $I(A: C | B) =0$, only when the regions have a disconnected minimal surface. Therefore, when $I(A: C | B) =0$, we have
\(
I(A:B) = I(A:C) = I(B:C) = 0
\)
Moreover, the conditions $I(A: C | B) =0$ and $I(B: C | D) =0$ imply that $I(C:D) = 0$ and $I(C: A \cup B) = 0$. Therefore, the Linden-Winter inequality is satisfied trivially.
\subsection{Cadney-Linden-Winter inequality}
We will proceed as we did in the previous section. $I(A: C | B) =0$ implies that  $I(B: C | A) =0$. For disjoint subsystems $\{ A,B,X_{1},X_{2},...,X_{n}\}$, $I(A: C | X_{i}) \geq 0$ follows from the strong subadditivity condition. By repeated application of the subadditivity condition, we get
\(
S(X_{1} \cup X_{2}... \cup X_{n}) \leq \sum_{i \leq n} S(X_{i})
\)
Combining the inequalities, we get
\(
S\left(X_{1} \cup X_{2}... \cup X_{n}\right)+(n-1) I(A \cup B: C) \leq \sum_{i=1}^{n} S\left(X_{i}\right)+\sum_{i=1}^{n} I\left(A: B | X_{i}\right)
\)

It is worth noting that the way the Linden-Winter and Cadney-Linden-Winter inequalities are {\em trivially} satisfied is identical to that in AdS. This suggests that the triviality of these relations may be an indication of holography, analogous to the monogamy of mutual information.
\subsection{Reflection Inequality}
\begin{figure}[t]
  \centering
  \begin{minipage}[b]{0.48\textwidth}
    \includegraphics[scale=0.41]{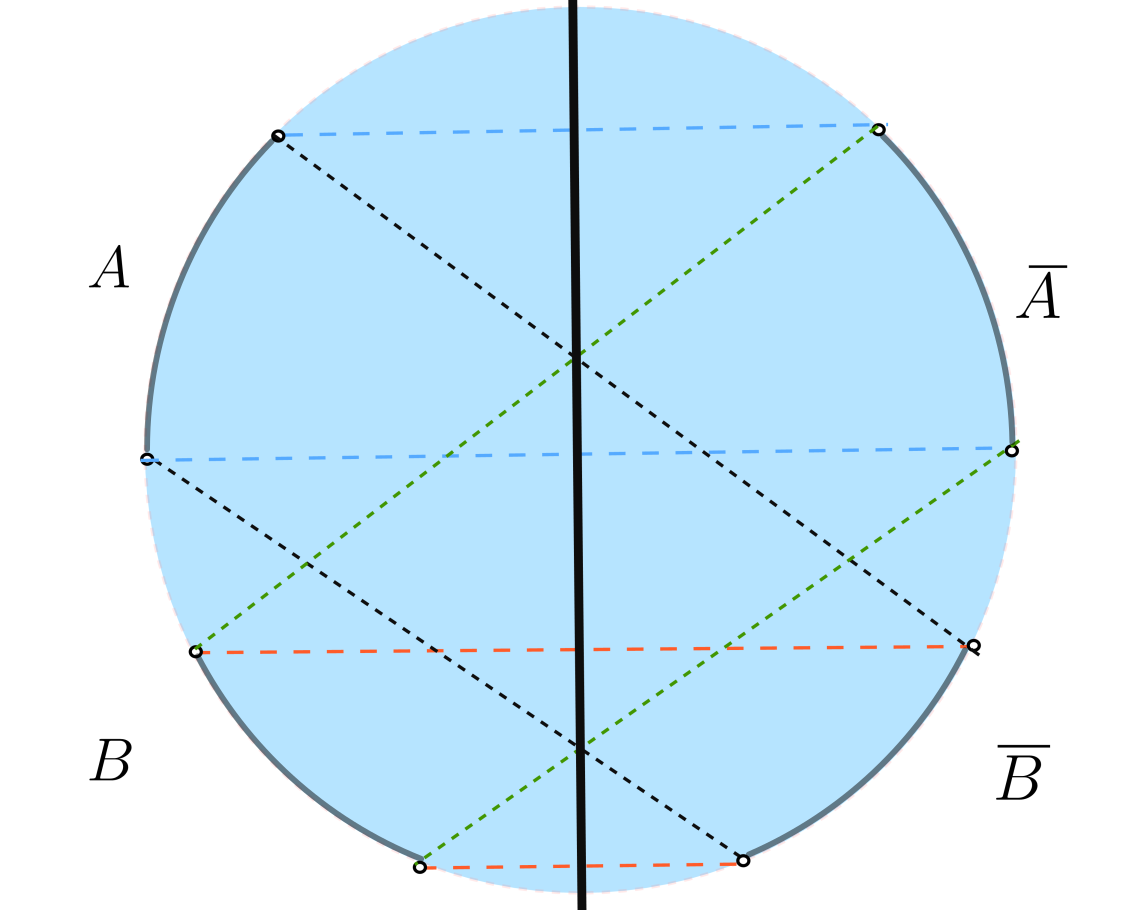}
\begin{center}
 (a)
\end{center}
  \end{minipage}
  \hfill
  \begin{minipage}[b]{0.48\textwidth}
    \includegraphics[scale=0.41]{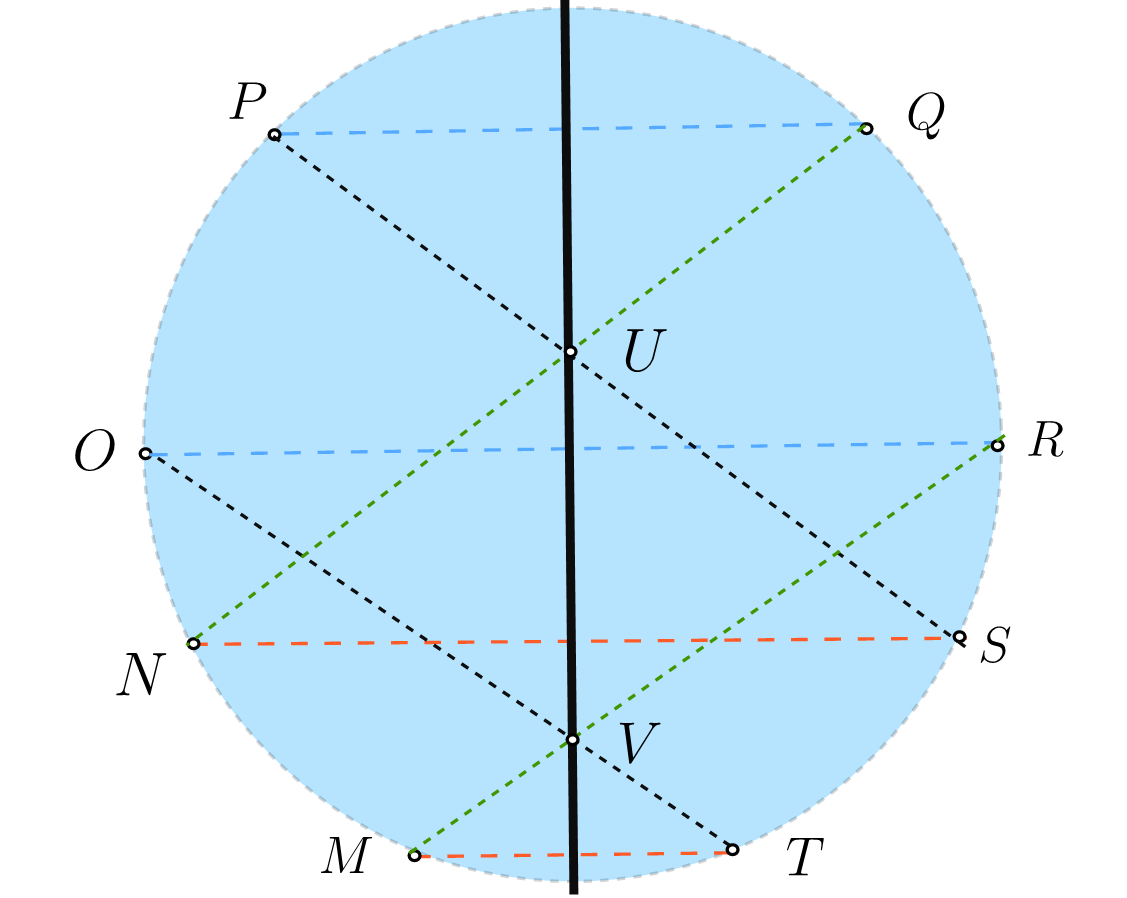}
\begin{center}
 (b)
\end{center}
  \end{minipage}
\caption{\textbf{a.} Consider two regions $A$, $B$ and their reflections along the solid black line. We will denote the reflected regions by $\bar{A}$ and $\bar{B}$. The minimal surfaces of $A \cup \bar{A}$ and $B \cup \bar{B}$ are represented by the blue and red dashed lines while the minimal surfaces of $A \cup \bar{B}$ and $B \cup \bar{A}$ are represented by the black and green dotted lines respectively. \textbf{b.} Representation of all the relevant surfaces using their endpoints on the cutoff surface}
\label{refineq}
\end{figure}
Consider two regions $A$, $B$ and their reflections about an axis as shown in figure \ref{refineq}(a).  From figure \ref{refineq}(b), we have 
\bea
\begin{aligned}
m_{A \cup \bar{B}} + m_{B \cup \bar{A}} &= (PS+ OT) + (NQ + MR) \\
&=(PU+US)+ (OV+VT) + (NU+UQ) +(MV+VR) \\
& =(PU+UQ) +(OV+VR) + (NU+US)+ (MV+VT)
\end{aligned}
\eea
By triangle inequality,
\bea
PU+UQ \geq PQ, \ OV+VR \geq OR, \ NU+US \geq NS, \ MV+VT \geq MT
\eea
Therefore,
\bea
\begin{aligned}
m_{A \cup \bar{B}} + m_{B \cup \bar{A}} &\geq (PQ+OR)+(NS+MT)\\
& = m_{A \cup \bar{A}}+m_{B \cup \bar{B}}
\end{aligned}
\eea
This give us the required inequality
\bea
S(A \cup \bar{A})+S(B \cup \bar{B}) \leq  S(A \cup \bar{B}) + S(B \cup \bar{A})
\eea
It is quite straightforward to see that a similar proof will go through in all the other relevant cases.
\section{Inequalities in Entanglement of Purification}

Entanglement of purification (EOP) is a measure that quantifies the entanglement of a mixed state. It has been conjectured that the boundary entanglement of purification is equal to the minimal cross-section area of the joint \textit{entanglement wedge}, which is a bulk region bounded by the boundary subsystems and their RT surfaces \cite{EOP1,EOP2}. We expect that this conjecture to hold in flat-space as well. We can check this proposal by explicitly verifying various inequalities EOP is expected to satisfy in general quantum mechanical systems.

\begin{figure}[t]
\centering
\includegraphics[scale=0.45]{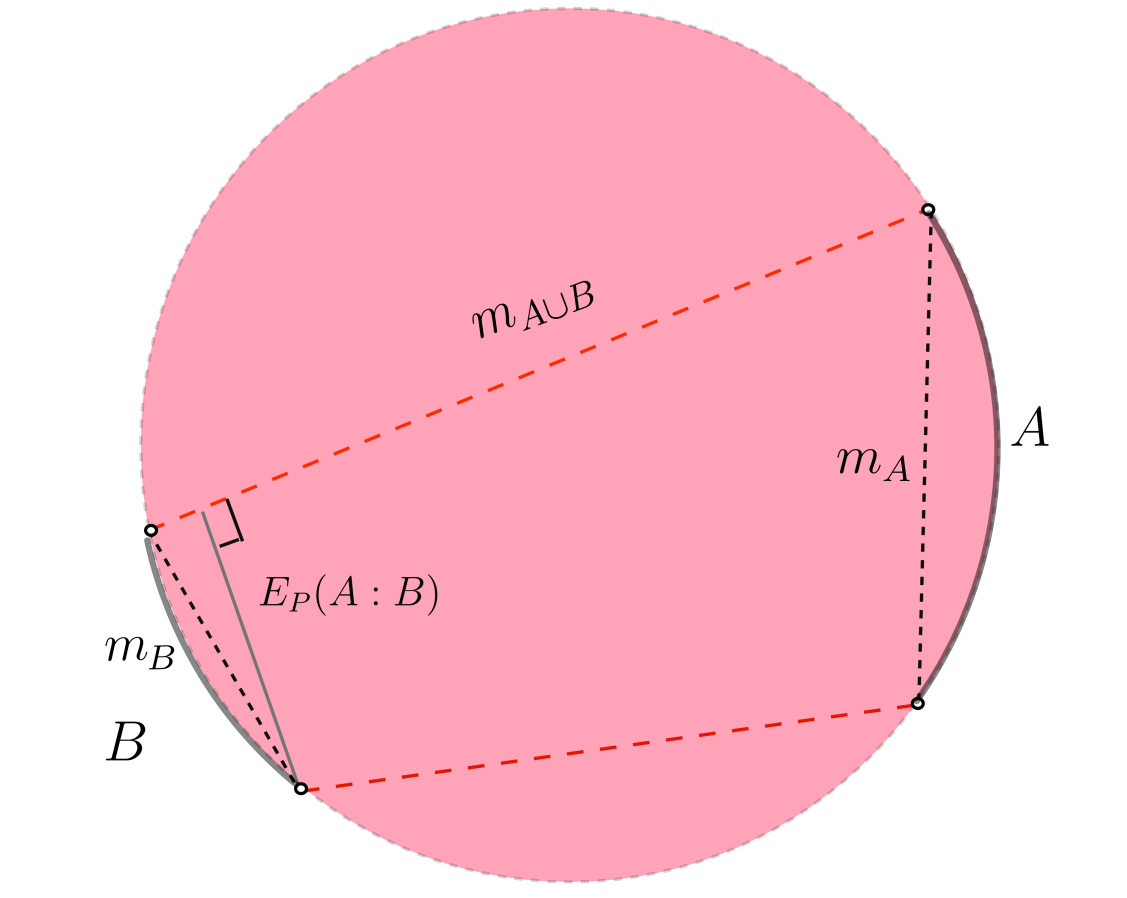}
\caption{For non adjacent regions $A$ and $B$, the minimal surfaces corresponding to $m_{A \cup B}$ is represented by the red dotted lines. Therefore, the entanglement wedge corresponding to these regions is the region bounded by the minimal surfaces of $A \cup B$, $A$ and $B$. The \textit{minimal cross-section} of this entanglement wedge is given by the length of the gray line. This line segment has the smallest length out of all the line segments ending on the minimal surfaces of $A \cup B$. }
\label{eop1}
\end{figure}

Let us state the explicit holographic proposal in general dimensions. Consider a bipartite system as shown in the figure \ref{eop1}. Let $\Sigma$ denote the minimal surface of the region $A \cup B$. The entanglement wedge corresponds to the region $A\cup B \cup \Sigma $. Let us look at the class of all surfaces that end on $\Sigma$ and is homologous to $A$. Let $X$ be the minimal area surface. Then, EOP of the regions $A$ and $B$ is given by
\(
E_{p}(A:B) = \frac{\text{Area}(X)}{4 G_{\mathrm{N}}}
\)
$E_{p}$ is non-zero only when the entanglement wedge is connected. The situation simplifies a lot in a $2+1$d Minkowski spacetime as the minimal surfaces are straight lines. Without loss of generality, let us assume that $S(A) \geq S(B)$. Therefore, $X$ corresponds to the perpendicular drawn from one of the endpoints of $B$ onto the minimal surface of $A \cup B$ in such a way that  $X$ lies inside the entanglement wedge (See figure \ref{eop1}).

EOP is expected to satisfy certain inequalities \cite{EOP2}
\begin{enumerate}
\item $E_{p}(A:B) \leq \text{min}(S(A),S(B))$ 
\item \textit{Monotonicity} :  $E_{p}(A:B \cup C) \geq E_{p}(A:B) $
\item  $E_{p}(A:B) \geq I(A:B)/2$ 
\item For a tripartite system $E_{p}(A:B  \cup C) \geq I(A:B)/2 + I(A:C)/2$ 
\item  For a tripartite pure system, $E_{p}$ is polygamous, i.e. $E_{p}(A:B )+E_{p}(A:C ) \geq E_{p}(A:B\cup C)$ 
\end{enumerate}
We expect our holographic proposal to satisfy these inequalities and we will show that this is indeed the case. 

\subsection{Proofs}

\textbf{1.} $E_{p}(A:B) \leq \text{min}(S(A),S(B))$
\begin{figure}[!tbp]
  \centering
  \begin{minipage}[b]{0.48\textwidth}
    \includegraphics[scale=0.41]{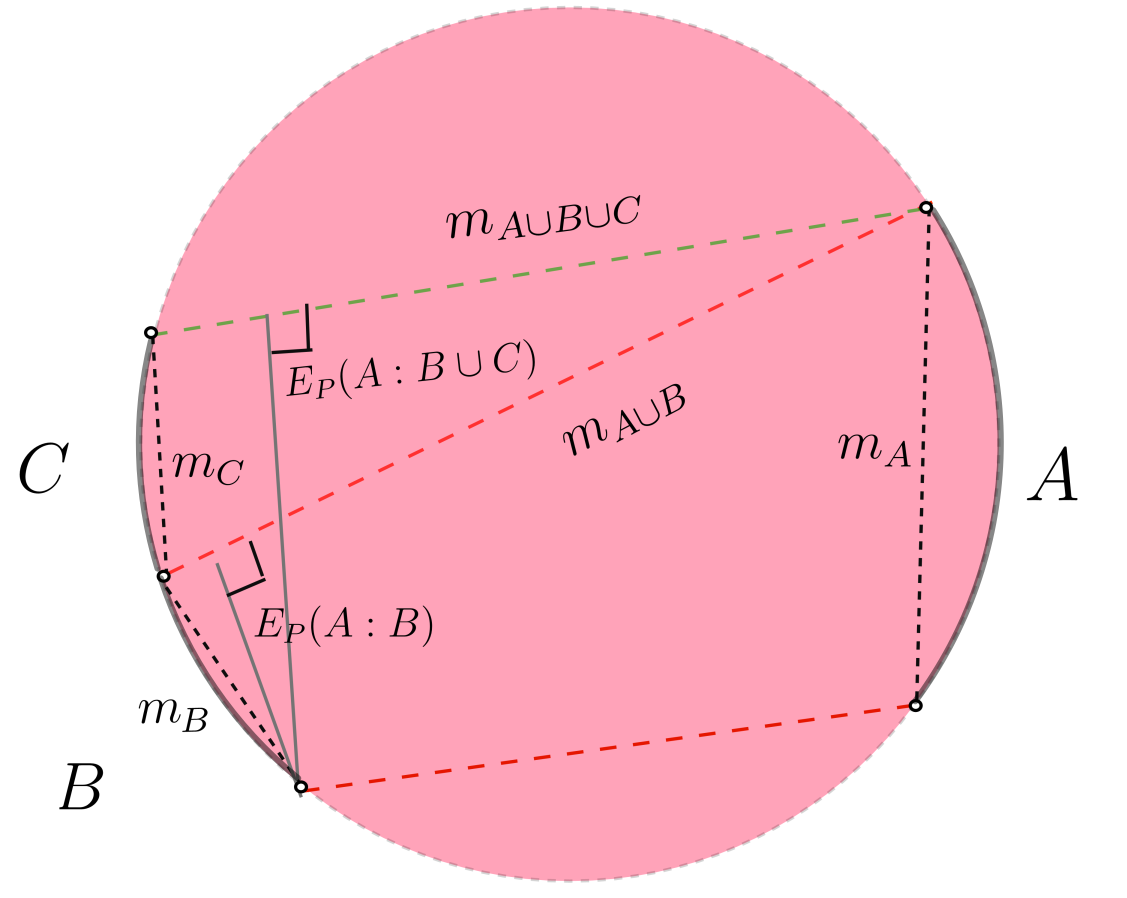}
\begin{center}
 (a)
\end{center}
  \end{minipage}
  \hfill
  \begin{minipage}[b]{0.48\textwidth}
    \includegraphics[scale=0.41]{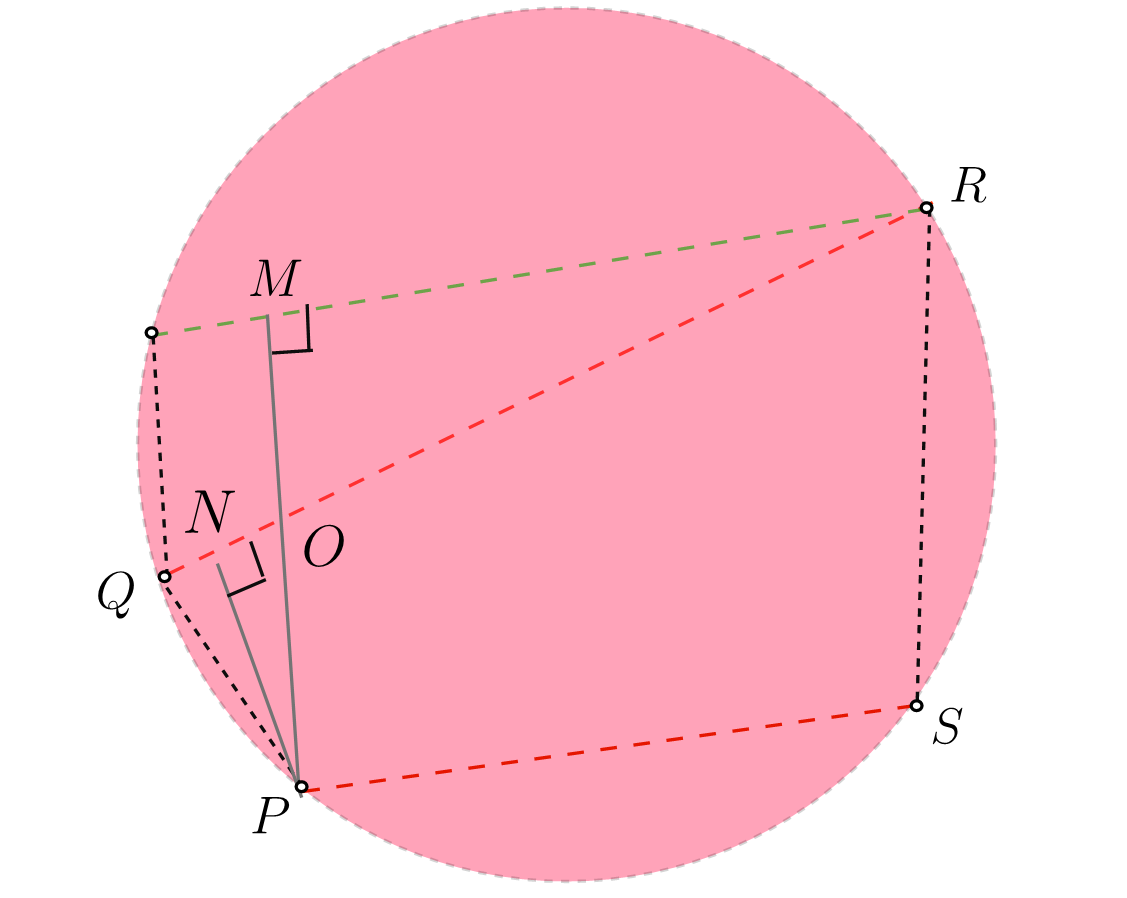}
\begin{center}
 (b)
\end{center}
  \end{minipage}
\caption{\textbf{a.} In the figure, the minimal surface of $A \cup B$ is given by the union of the red dotted lines while the minimal surface of $A \cup B \cup C$ is given by the union of the green line and the bottom red line. The EOP of the bipartite/tripartite system is given by the length of the gray line. \textbf{b.} Representation of the relevant surfaces and their endpoints.}
\label{eop2}
\end{figure}

Let us assume that $S(A) \geq S(B)$. We can see from figure \ref{eop1} that Pythagoras theorem holds for $X$ and the minimal surface of the region $B$. Therefore,
\(
 \text{length}(X) \leq m_{B} \leq m_{A} 
\)
This implies that
\(
E_{p}(A:B) \leq \text{min}(S(A),S(B))
\)

\textbf{2.} $E_{p}(A:B \cup C) > E_{p}(A:B)$

Let $X$ and $Y$ be the minimal cross-sections of the tripartite and bipartite systems respectively. Let us assume that $X$ intersects the minimal surface of $A \cup B$ at $O$, as shown in figure \ref{eop2}(a). Therefore, we have 
\(
PM = PO+ OM  \geq PO  \ \ \ \ \ 
\)
Let $P$ and $N$ be the endpoints of the $Y$. Since $Y$ is the minimal cross-section, $PN$ will be less than the length of any curve connecting the minimal surfaces of $A \cup B$. Therefore
\(
PM \geq PO \geq PN \implies  E_{p}(A:B \cup C) \geq  E_{p}(A:B)
\)

\textbf{3.}  $E_{p}(A:B) \geq I(A:B)/2$

Let us look at figure \ref{eop2}(b) and consider the following sum:
\(
\begin{aligned}
2 PN + QR+ PS = 2 PN + QN+ NR +PS = (PN+ QN)+ (PN+NR+PS)
\end{aligned}
\)
We have
\(
PQ \leq QN+ NP  \ \ \ \ 
\)
and
\(
(PN+NR)+PS \geq PR+PS \geq RS
\)
where we have used triangle inequality twice to obtain the inequality. Therefore, we have
\(
\begin{aligned}
 (PN+ QN)+ (PN+NR+PS) \geq PQ+RS \implies  2 PN \geq PQ+RS -(QR+ PS)
\end{aligned}
\)
Writing this in terms of the entanglement entropy, we get 
\(
E_{p}(A:B) \geq I(A:B)/2
\)

\textbf{4.}  $E_{p}(A:B  \cup C) \geq I(A:B)/2 + I(A:C)/2$ 
\begin{figure}[t]
  \centering
  \begin{minipage}[b]{0.48\textwidth}
    \includegraphics[scale=0.41]{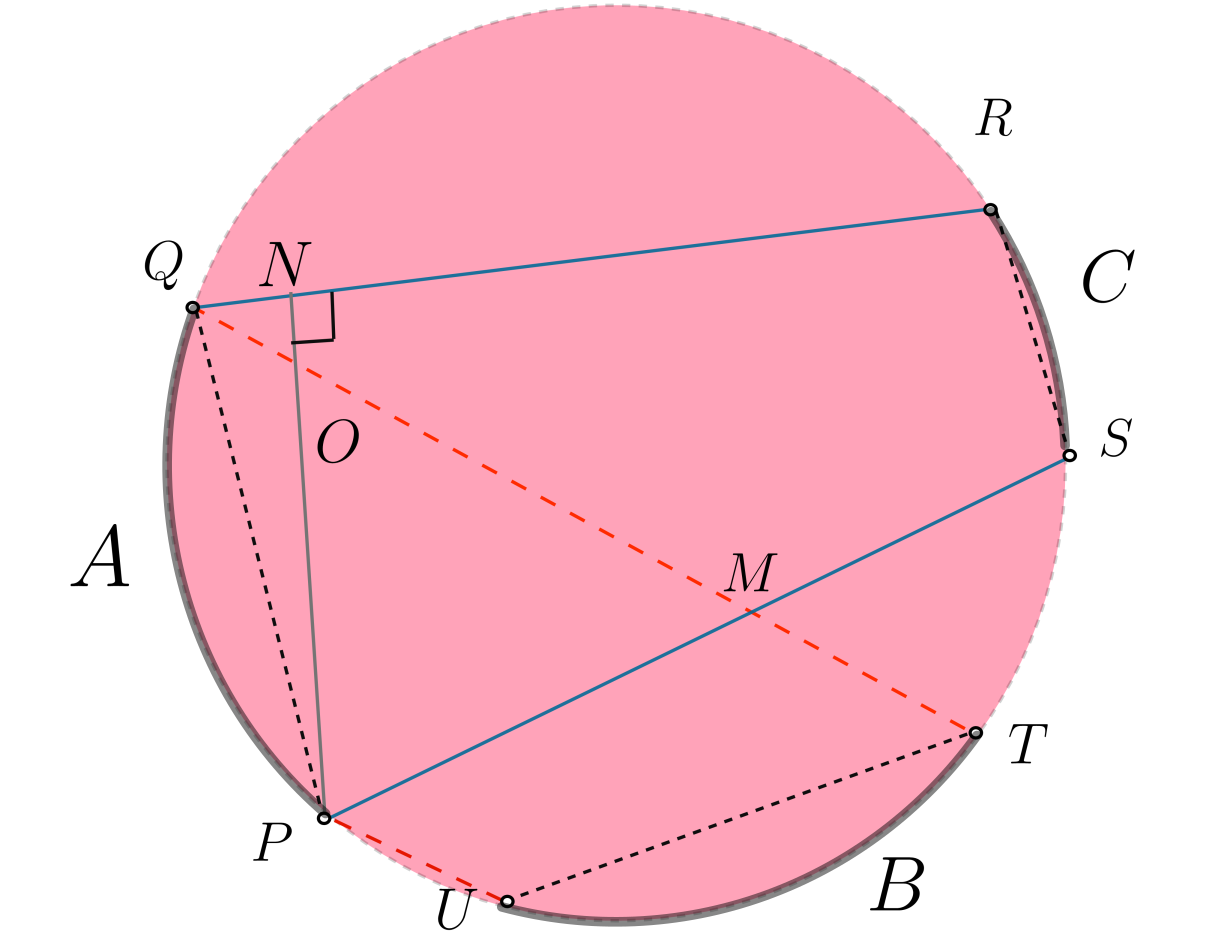}
\begin{center}
 (a)
\end{center}
  \end{minipage}
  \hfill
  \begin{minipage}[b]{0.48\textwidth}
    \includegraphics[scale=0.41]{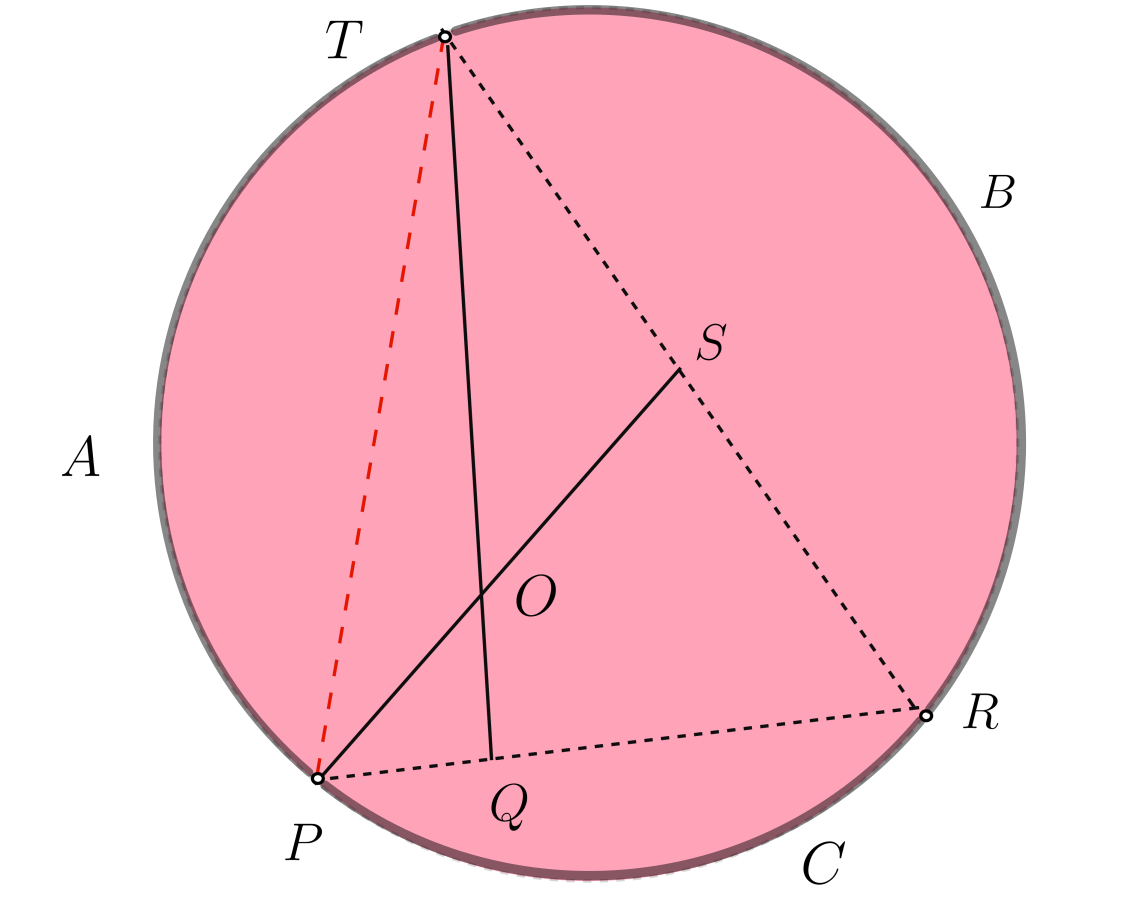}
\begin{center}
 (b)
\end{center}
  \end{minipage}
\caption{\textbf{a.} Consider three non-adjacent regions $A$, $B$ and $C$ (whose minimal surfaces are shown by black dotted line). The minimal surface of $A \cup B$ is given by union of the red lines. Similarly, the minimal surface of $A \cup C$ is given by the union of the blue lines. The EOP is given by the length of the gray line $PN$. \textbf{b.} The figure shows a tripartite pure system. The length of the minimal cross-section surfaces of $A \cup B$ and $A \cup C$ are given by $TQ$ and $PS$ respectively. In this case, $E_{p}(A:B \cup C)$ is given by the minimal surface of the region A. }
\label{eop3}
\end{figure}

Consider a tripartite system as shown on figure \ref{eop3}(a.). Let us focus on the quantity 
\(
\begin{aligned}
&4 G_{\mathrm{N}} (2 E_{p}(A:B \cup C) + S(A \cup B)+S(A \cup C)) \\
&= 2 PN + (QT+PU) +(QR+PS) \\
&=PN+ (PO+ON)+(QO+OM+MT)+PU+(QN+NR)+(PM+MS)\\
& =(PN+QN)+(PO+QO)+(ON+OM+MS+NR)+(PM+MT+PU)
\end{aligned}
\)
Now we will analyze the terms in parenthesis in the last equation separately. We have 
\(
PN+QN   \geq PQ,  \ \ \ \ \ PO+QO   \geq PQ
\)
and (by repeated usage of triangle inequality), 
\(
(PM+MT)+PU \geq PT +PU\geq UT
\)
Similarly
\(
ON+(OM+MS)+NR\geq (ON+OS)+NR \geq SN+NR \geq RS
\)
Therefore, we have 
\(
\begin{aligned}
&4 G_{\mathrm{N}} (2 E_{p}(A:B \cup C) + S(A \cup B)+S(A \cup C)) \\
& =(PN+QN)+(PO+QO)+(ON+OM+MS+NR)+(PM+MT+PU)\\
&\geq 2 PQ +UT+RS =4 G_{\mathrm{N}} (2 S(A)+ S( B)+S( C))
\end{aligned}
\)
Rearranging the terms, we get the inequality
\(
E_{p}(A:B  \cup C) \geq I(A:B)/2 + I(A:C)/2
\)

\textbf{5.}  $E_{p}(A:B )+E_{p}(A:C ) \geq E_{p}(A:B\cup C)$

Consider a pure tripartite system as shown in figure \ref{eop3}(b.). We can see that the EOP of the system will be given by 
\(
E_{p}(A:B\cup C) = PT/4 G_{\mathrm{N}}
\)
Therefore, let us consider the LHS of the inequality
\(
\begin{aligned}
E_{p}(A:B )+E_{p}(A:C )& = (PS+QT)/4 G_{\mathrm{N}}\\
& = (PO+OS+QO+OT)/4 G_{\mathrm{N}}\\
&\geq (PO+OT)/4 G_{\mathrm{N}}
\end{aligned}
\)
Using
\(
PO+OT \geq PT
\)
we get,
\(
\begin{aligned}
E_{p}(A:B )+E_{p}(A:C )& = (PS+QT)/4 G_{\mathrm{N}}\\
&\geq (PO+OT)/4 G_{\mathrm{N}} \\
&\geq (PT)/4 G_{\mathrm{N}}  =E_{p}(A:B\cup C)
\end{aligned}
\)

\section{Punchline}

The proofs in this paper were trivial, just as they were in similar discussions in AdS\footnote{In fact our proofs were simpler even than in AdS. We found that triangle inequality is in some sense the ultimate entanglement inequality. It is coneceivable that these are the simplest proofs of entanglement inequalities, possible. Note that in non-holographic systems these proofs are often quite technical, eg. \cite{Lieb}.}, but the conclusions do not seem as widely appreciated as they should be. They suggest that even in the bulk of flat pace, subregions of screens capture some form of holographic entanglement entropy. 

The immediate reason we feel this is worth emphasis is due to the current debate on the conventional tent-shaped Page curve in long range gravity. It has been suggested that in holographic theories, the only meaningful Page curve is the flat Page curve \cite{Alok, Andreas}. There is no doubt that a flat Page curve can be found in holography, it means that all the information is always present at infinity \cite{DeWitt}. Note however that the availability of all the information at infinity says {\em nothing} about whether there is any information in the bulk. The point of holography is that information is in fact encoded {\em redundantly} in the radial direction as well. This is worth emphasis, because we feel there is an implicit conflation of the following two statements in many of these discussions:
\begin{itemize}
\item All information is always available at infinity in gravity,  
\vspace{0.1in}
\item There is no information available in the bulk in gravity
\end{itemize}
We would of course agree with the former statement in holographic theories, but not the latter.

The ideas presented in \cite{Jude, Maldacena} suggest that screens at  various epochs of Hawking radiation  do provide a useful characterization of information in the {\em bulk}. The existence of Page phase transitions of quantum extremal surfaces defined with respect to screens, was viewed as evidence for this. The existence of such phase transitions, between extremal surfaces associated to an information paradox and those that resolve it, has now been observed in multiple distinct contexts for surfaces anchored to screens \cite{Thorlacius, Iizuka1, Iizuka2, Jude, Maldacena, Critical, Kausik}. These cases cover different kinds of gravitating baths, and yet the island physics remains intact. We would specifically like to point out the case considered in \cite{Critical} where the cosmological redshifting on a braneworld played a crucial and intuitive role in resolving the paradox\footnote{Incidentally, this is the first example of a cosmological island that seems to have appeared in the literature.}. 

And yet, a committed critic may view these simply as phase transitions in extremal surfaces, without any particular meaning as far as quantum information is considered\footnote{cf., footnote \ref{foot}.}. The results of our present paper make it harder to do so. On top of the strong subadditivity inequalities we proved in \cite{ACD, Jude}, in this paper we have showed that areas of extremal surfaces on subregions on screens in Minkowski space satisfy a long list of properties that are satisfied by entanglement entropy. This observation is eminently sensible, in a world where the conventional bulk Page curve is meaningful. On the contrary, claims that the bulk Page curve in flat space is meaningless and merely a geometric curiosity about extremal surfaces, will struggle to explain these observations.

We believe that these results are quite strong already as statements in Minkowski space, because there was {\em no reason} for them to be satisfied other than due to some form of holography on the screen. Let us emphasize that these are {\em not} statements about codimension-1 (spatial) subregions of Minkowski where we would expect ordinary quantum field theoretic entanglement entropy relations to hold, when gravity decouples. Instead, these are statements about subregions {\em on screens} in Minkowski space and extremal surfaces anchored to them. Non-gravitational field theory in Minkowski space can {\em not} immediately explain this - for the same reason that Ryu-Takayangi formula in a fixed AdS space does not follow via QFT in the bulk arguments. Therefore these observations should be viewed as holding semi-classically in the weak gravity limit when $G_N \rightarrow 0$. This is in line with the results in \cite{Kausik} where it was observed that moving the anchoring surface of the RT surfaces to a location where gravity decouples, does {\em not} qualitatively change the phenomenology of RT surfaces and their phase transitions\footnote{Note that one needs a set up where the bulk and the bath remain intact as gravity decouples at the anchor, in order to study this. For Schwarzschild black hole in flat space, if one takes the anchor to the asymptotic region, the bath also vanishes and the problem becomes uninteresting. This was one of the major motivations behind \cite{Kausik}.}.

Euclidean versions of our results are in fact consistent with the arguments in \cite{Lew}. The replica method  based {\em bulk} arguments there lead to areas of extremal surfaces. While the context of the standard Ryu-Takayanagi prescription is when these extremal surfaces are anchored to an AdS boundary, the bulk argument is not sensitive to that. If one simply views the cut-off as the location of the dual replica argument, even in flat space we will find the requisite extremal surfaces anchored to subregions. This again suggests that one should be able to define an entanglement entropy for subregions on the cut-off/screen. 

An incomplete list of recent works on the island program are collected in \cite{summary}.


\section{Acknowledgments}

We thank the organizers and participants of ISM2021, Roorkee, for a stimulating discussions on related questions.

\appendix
\section{Another Proof for Strong Subadditivity I}
\label{subaddappend}
\begin{figure}
  \centering
  \begin{minipage}[b]{0.48\textwidth}
    \includegraphics[scale=0.41]{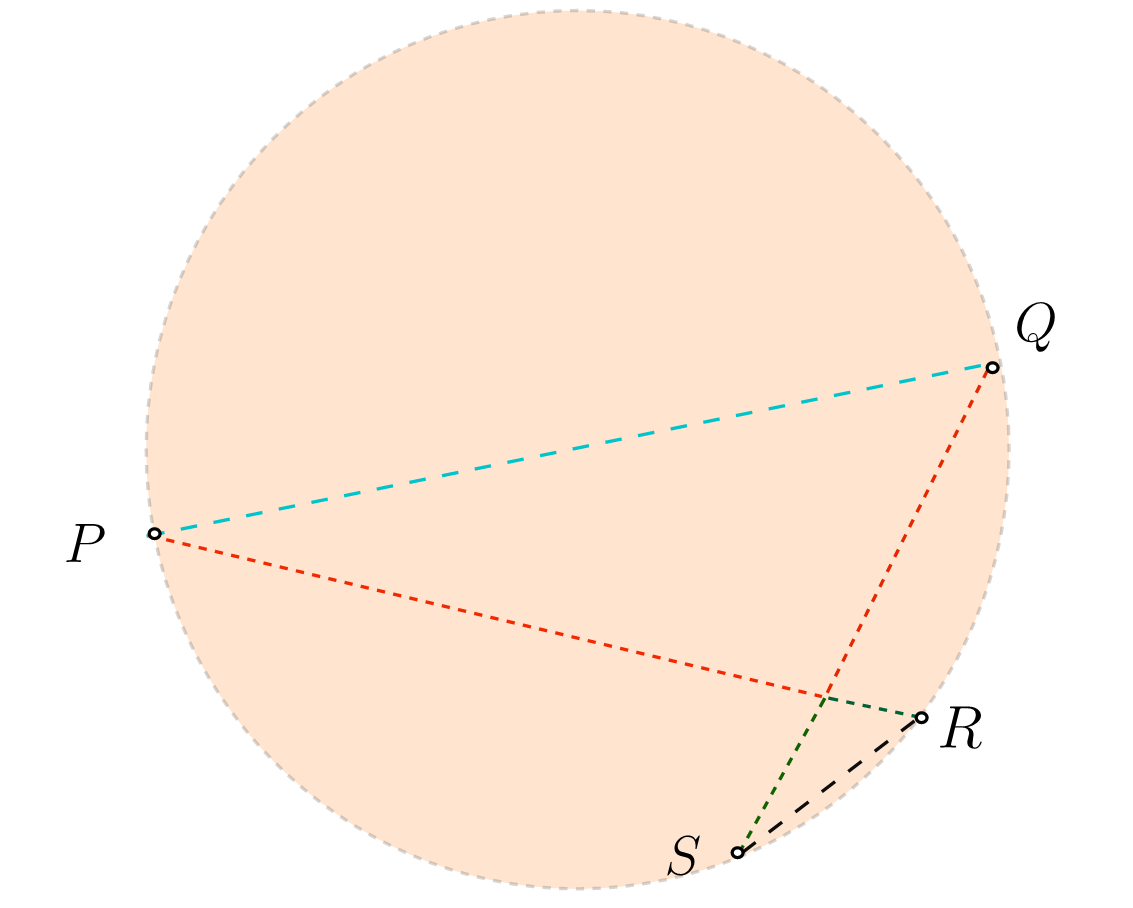}
\begin{center}
 (a)
\end{center}
  \end{minipage}
  \hfill
  \begin{minipage}[b]{0.48\textwidth}
    \includegraphics[scale=0.41]{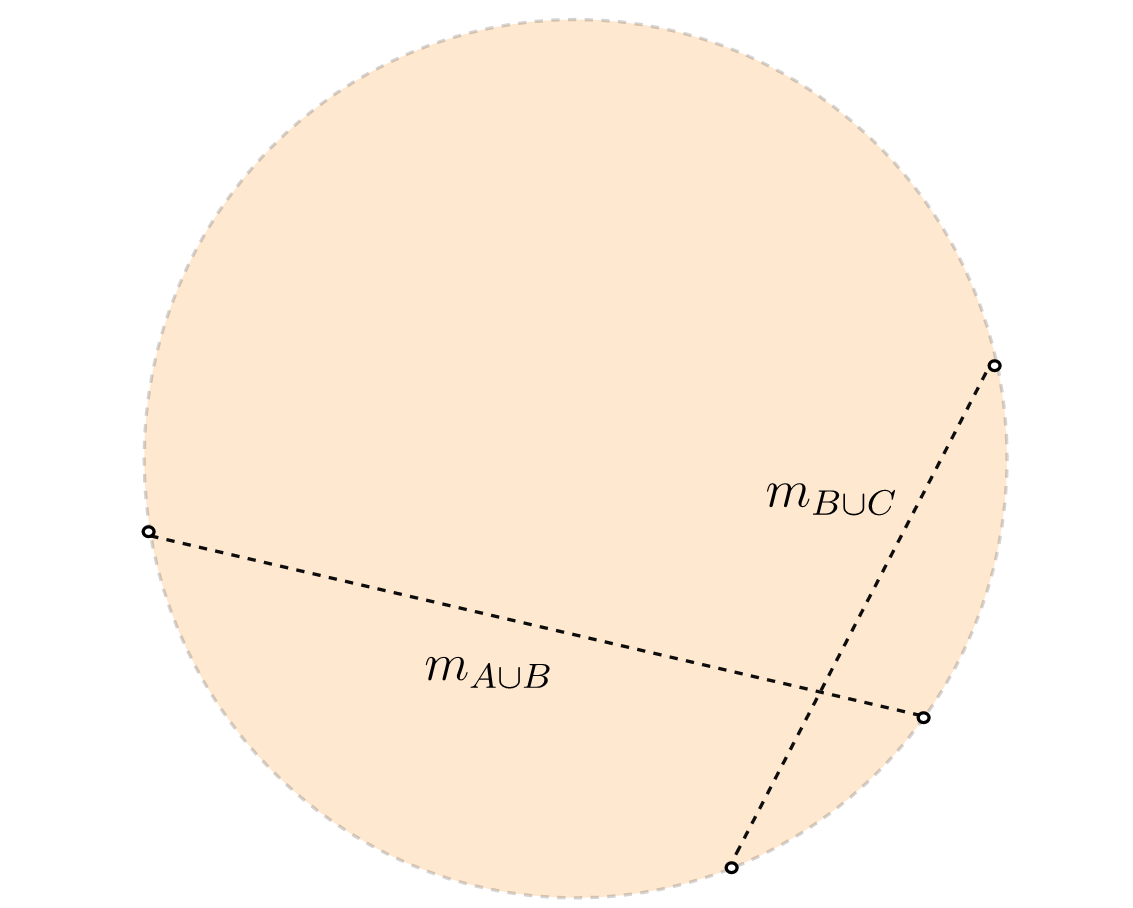}
\begin{center}
 (b)
\end{center}
  \end{minipage}
\caption{\textbf{a.} Consider three non-adjacent regions $A$, $B$ and $C$ as in figure \ref{strongsubfig1}(a). The minimal surfaces of $A \cup B$ and $A \cup C$ are given by the black dotted lines. \textbf{b.} We perform a local \textit{surgery} at the point of intersection. The length of the red curve will be more than that of the blue line as the line is the minimal surface connecting $P$ and $Q$. Similarly, the green curve will have more length than the black line.}
\label{append2}
\end{figure}

We provide another proof for strong subadditivity I by performing a \textit{surgery} at intersection of minimal surfaces. This is just a fancy word for the following. Consider three adjacent disjoint regions $A$, $B$ and $C$ as in figure \ref{strongsubfig1}(a). The minimal surfaces of $A \cup B$ and $B \cup C$ are shown in figure \ref{append2}(a).
Now consider the point where the two lines intersect. We can partition the minimal surfaces as shown in figure \ref{strongsubfig1}(b). The union of the red line segments corresponds to a curve that starts at the point $P$ and ends at $Q$. Therefore, the length of this curve will be larger than $m_{A \cup B \cup C}$.

Similarly, the green line segments correspond to a curve joining $R$ and $S$ and will be longer than the minimal surface of $B$. This gives us
\(
m_{A \cup B}+m_{B \cup C} \geq m_{A \cup B \cup C} +m_{B}.
\)
This implies that the strong subadditivity I holds. 

It is very straightforward to see that this technique can be used to prove all of the more cumbersome inequalities in this paper. In fact, it also suggests a reverse process: we could use this as a tool for {\em generating} inequalities that can then be candidate inequalities in generic or holographic quantum systems.

A careful inspection shows that the proof relied on a particular feature of the underlying space - the entanglement wedge of a region $A$ is the subset of the entanglement wedge of the region $A \cup B$. Flat spacetime trivially satisfies this condition. A similar discussion can be found in \cite{Rangamani} in the context of AdS/CFT correspondence.

\end{document}